\documentclass{JINST}

\usepackage{graphicx}
\usepackage{subfigure}
\usepackage{amssymb,amsmath}

\title{Absorbing systematic effects to obtain a better background model in a search for new physics}

\author{S. Caron$^a$, G. Cowan$^b$, E. Gross$^c$, S. Horner$^a$\thanks{Corresponding
author.}~ and J. E. Sundermann$^a$\\
\llap{$^a$}Physikalisches Institut,\\
  Albert-Ludwigs-Universit\"at Freiburg,\\ Hermann-Herder-Stra\ss e 3, 79104 Freiburg i.~Br., Germany\\
\llap{$^b$}Royal Holloway, \\ University of London,\\ Egham, Surrey TW20 0EX, UK.\\
\llap{$^c$}Weizmann Institute of Science,\\
  Rehovot 76100, Israel\\ \\
  E-mail: \email{stephan.horner@physik.uni-freiburg.de}\\}

\abstract{This paper presents a novel approach to estimate the Standard Model backgrounds based on modifying Monte Carlo predictions within their systematic
uncertainties. The improved background model is obtained by altering the original predictions with successively more complex correction functions in
signal-free control selections.  
Statistical tests indicate when sufficient compatibility with data is reached. In this way, systematic effects are absorbed into the new 
background model. The same correction is then applied on the Monte Carlo prediction in the signal region.
Comparing this method to other background estimation techniques shows improvements with respect to statistical and systematic uncertainties. The proposed
method can also be applied in other fields beyond high energy physics.}

\keywords{Analysis and statistical methods, Data processing methods}

\begin{document}
\section{Introduction}

The way to discover new physics beyond the Standard Model (SM) is to measure a significant deviation from the 
SM prediction in a signal region, that is a region of phase space where new physics is expected to appear.
It is therefore essential that one can have utmost confidence in an estimate on this SM prediction in order to avoid false discoveries and 
overlooked signals. State-of-the-art Monte Carlo (MC) generators yield such estimates by modelling the relevant physics processes. However, systematic effects due to an imperfect detector and shortcomings in the underlying models of the MC generators lead
to an insufficient description of the data. A way to verify and improve the validity of the MC prediction is to compare it with data in a
signal-free control region in phase space. 

Usually, measurements in the control region have to satisfy certain requirements. An observable of interest $x$ needs to have similar physical meanings and dependencies on systematic effects in both signal and control regions. Remaining differences of these lead to systematic uncertainties which need to be estimated, but are not covered in this paper.

\begin{figure}[htbp]
  \begin{center}
    \includegraphics[width=0.70\textwidth]{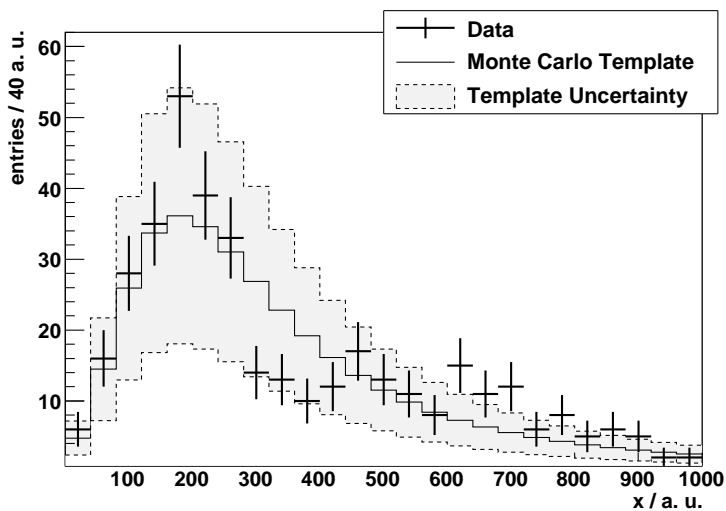}
  \end{center}
  \caption{ Example of a data distribution and corresponding Monte Carlo estimate with its systematic uncertainty in a signal-free control region.
  }\label{fig:IntroPlot}
\end{figure}

Figure  \ref{fig:IntroPlot} shows an example of a data distribution measured in a signal-free control region with its corresponding MC prediction.

\noindent To keep things simple it is here assumed that all data originate from a single process and that a sufficiently large MC sample is available resulting in a smooth curve with negligible statistical uncertainty. Within the large systematic uncertainty data and Monte Carlo estimate are in agreement. However,
the data are not fully described by the central MC prediction.

In the following, a method is presented which incorporates this systematic deviation between the data and the original model into an improved background
model.

\section{Concept}

The proposal of this paper is to reweight the MC estimate by multiplying it with an appropriate correction function. The correction function depends on a set of 
adjustable parameters which are determined by fitting the modified estimate to the data in the control region. Then the same function is to be applied on the corresponding template in the signal region. In general, the statistical uncertainty grows with the number of parameters. Since this method
 starts with the original MC expectation, the number of parameters needed for the correction is in general smaller than in other fitting approaches, where one
fits a function or splines to the data. 

Another advantage of this method is that the templates in control and signal region need not have identical shapes, only the systematic effects have to influence them in a similar way. This procedure can be easily generalized to the carrying out of a combined fit of several data distributions arising from multiple sources.

In order to avoid mistaking a real signal in the control region as a systematic effect of the detector or the physics modelling, the Monte Carlo templates can be varied according to known sources for systematic deviations (energy scales, efficiencies, PDF uncertainties, etc.) before the
fit is carried out. These variations yield constraints on the expected systematic deviation of the original template when there is no signal present. Only if the data is compatible with those the fit is followed through (cf.~the template uncertainty in figure \ref{fig:IntroPlot}).

Obviously, the crucial point lies in choosing the right correction functions which, together with the MC template, constitute the ``best model'' to describe the data.

\section{Determination of the best model}

Were there no systematics present, plain Monte Carlo estimates would suffice in describing the different contribution to the data. Thus, those templates form the natural starting point to determine the best background description. One modifies them with correction functions of increasing complexity, that is, with an increasing number of free parameters, thereby allowing greater flexibility for the adjustment to data until the goodness-of-fit reaches a certain level. It is useful to take functions forming a complete basis set
such as certain kinds of polynomials.

The starting model, the unaltered Monte Carlo template, shall be labeled zeroth-order model. The mean number of entries in each bin ${\vec{\nu}} = (\nu_1, \ldots, \nu_N)$ predicted by this
model constitute the template histogram. Assuming that the data is independently Poisson distributed, the probability to observe the data $\vec{n} = (n_1, \ldots, n_N)$ is

\begin{equation}
\label{eq:poisson}
P(\vec{n}; \vec{\nu}) = \prod_{i=1}^N \frac{\nu_i^{n_i}}{n_i!} e^{-\nu_i} \;.
\end{equation}

\noindent To quantify the level of compatibility between $\vec{n}$ and $\vec{\nu}$ one could compute Pearson's chi-square statistic,

\begin{equation}
\label{eq:pearson}
\chi^2_{\rm P} = \sum_{i=1}^N \frac{ (n_i - \nu_i)^2 }{ \nu_i } \;.
\end{equation}

\noindent An almost equivalent statistic is based on the likelihood ratio,

\begin{equation}
\label{eq:lr}
\lambda{(\vec{\nu})} = \frac{L(\vec{\nu})}{L(\hat{\vec{\nu}})},
\end{equation}

\noindent where $L(\vec{\nu}) = P(\vec{n};\vec{\nu})$ is the likelihood
of the hypothesized model $\vec{\nu}$, and $\hat{\vec{\nu}}$ is
the maximum likelihood (ML) estimator for $\vec{\nu}$, i.e., the
values of $\nu_1, \ldots, \nu_N$ which maximize the likelihood.
By setting the derivative of $L(\vec{\nu})$ equal to zero and solving
one easily finds 

\begin{equation}
\label{eq:mle}
\hat{\nu}_i = n_i
\end{equation}

\noindent for all $i$.

If the model $\vec{\nu}$ is correct, then Wilks' theorem \cite{Wilks}
states that the distribution of the statistic

\begin{equation}
\label{eq:qabs}
q_{\vec{\nu}} = - 2 \ln \lambda(\vec{\nu}) = 
2 \sum_{i=1}^N \left( n_i \ln \frac{n_i}{\nu_i} + \nu_i - n_i \right) 
\end{equation}

\noindent approaches a chi-square distribution for a sufficiently large data sample.\footnote{In computing
$q_{\vec{\nu}}$, the logarithmic term should be skipped if $n_i = 0$.} The number of degrees of freedom is the 
difference in the number of free parameters, often called the parameters of interest, of the two likelihood functions
in equation (\ref{eq:lr}). Here, the likelihood function in the numerator has no free parameters whereas in the 
denominator the number of free parameters is equal to the number of bins $N$ since the mean values $\nu_i$ are 
independently adjusted to the data values $n_i$ for each bin. 

In fact in many practical examples the chi-square approximation is
extremely good even for moderate samples, e.g., $n_i$ roughly a half
dozen or more.  Details on the regularity conditions required for
Wilks' theorem to be valid are discussed in standard texts such as
\cite{Kendall2}.  Pearson's $\chi^2_{\rm P}$ and the statistic
$q_{\vec{\nu}}$ are for the present example very similar; here
$q_{\vec{\nu}}$ will be used.

For either goodness-of-fit statistic,  $\chi^2_{\rm P}$ or $q_{\vec{\nu}}$,
one would quantify the compatibility between data and model by
giving the $p$-value.  This is the probability, under assumption
of the model $\vec{\nu}$, to obtain a value of the statistic
greater than or equal to that found with the actual data.  That is,

\begin{equation}
\label{eq:pval}
p = \int_{q_{\vec{\nu}, {\rm obs}}}^{\infty} f_{\chi^2} (z; N) \, dz \;,
\end{equation}

\noindent where

\begin{equation}
\label{eq:chi2dist}
f_{\chi^2} (z; N) = \frac{1}{2^{N/2}\Gamma(N/2)} z^{N/2 - 1}e^{-z/2} 
\end{equation}

\noindent is the chi-square distribution for $N$ degrees of freedom,
and $\Gamma$ is the Euler gamma function.

If the compatibility between the data and the zeroth-order model turns out to be unsatisfactory, one tries to improve 
the level of agreement by multiplying the template with a suitable correction function $s(x; \vec{\theta})$ as suggested 
above. The modified prediction for the mean number of entries in the $i$th bin is then
\begin{equation}
\label{eq:nuimod}
\nu_i \rightarrow \nu_i s(x_i; \vec{\theta}) \;,
\end{equation}
where $x_i$ is the value of the abscissa variable in the centre of the $i$th bin and $\vec{\theta}$ stands for the
set of \textit{M} adjustable parameters $\vec{\theta} = (\theta_1, \ldots, \theta_M)$ of the function. One can use the same 
ratio as in equation (\ref{eq:lr})  to assess the goodness-of-fit with only its numerator replaced by the likelihood of the 
modified prediction. Consequently, the number of degrees of freedom is reduced to $N-M$. This test can be applied on correction 
functions with an increasing number of adjustable parameters until the $p$-value exceeds a given threshold, say 0.1 or 0.2. 
Alternatively, one could pick the model with the highest $p$-value. The corresponding function together with the template would
then constitute the model with the smallest complexity to be compatible with the data, using a particular set of correction functions. 

One can, however, check whether the next more general model would significantly improve the data description by using another, 
closely related, test statistic which is based on the ratio of the likelihoods of the two models with $m$ and $m+1$ parameters 
respectively:

\begin{equation}
\label{eq:qrel}
q_{m,m+1} = - 2 \ln \frac{L(\hat{\vec{\theta}}^{(m)})}{L(\hat{\vec{\theta}}^{(m+1)})} \;.
\end{equation}

\noindent Under the assumption that the more restrictive model in the numerator
is correct and providing the data sample is not too small, $q_{m,m+1}$
will follow a chi-square distribution for one degree of
freedom. For this to be true the model in the numerator has to be a sub-model of the one in the demoninator, the same being true for $q_{\vec{\nu}}$ above.

Hence, the proposed strategy to determine an improved model for a certain background is twofold. First, an absolute goodness-of-fit using $q_{\vec{\nu}}$ establishes
the minimal number of needed parameters for a sufficient data description. Then a relative test via $q_{m,m+1}$ examines whether or not a substantially better 
prediction can be obtained by moving on to a more general model.

It has to be noted that the choice of a set of correction functions is obviously not unique. Different situations may call for different sets and a close
inspection beforehand might hint at a certain choice. In this chapter, two different basis sets are discussed: Ordinary 
and Bernstein polynomials.

\subsection{Ordinary polynomials}

\begin{figure}[tbp]
\begin{center}

\subfigure{
\includegraphics[width= 
0.45\textwidth]{./figures/IntroPlot.eps}
}
\subfigure{
\includegraphics[width= 
0.45\textwidth]{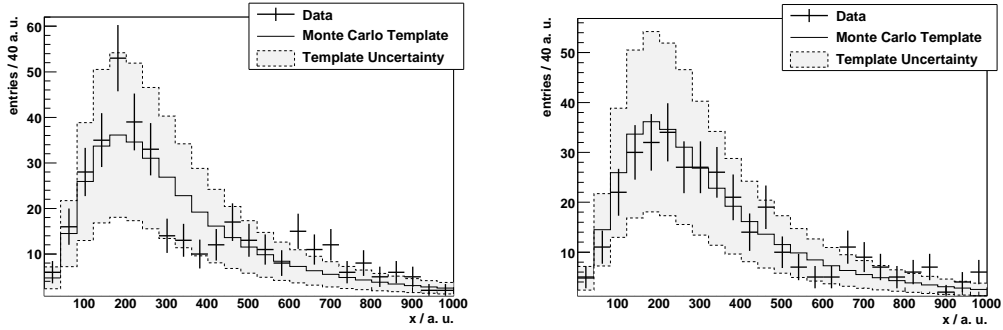}
}
\caption{ Two scenarios for measurements in a control region having the same Monte Carlo prediction.
}\label{fig:TwoScenarios}
\end{center}
\end{figure}

Consider two possible scenarios for measurements in a control region shown in figure \ref{fig:TwoScenarios}. Both measurements are
compatible with the uncertainty of their respective predictions. In the left plot, however, the data strongly deviate from the central prediction which
hints at substantial systematic effects being present in that scenario. This assumption is further supported by a $p(q_{\vec{\nu}})$ value of only about $0.3\%$. On the other hand, the deviations in the right plot seem compatible with statistical fluctations as is reflected by a $p(q_{\vec{\nu}})$ value of about $56\%$. 
In both cases, the data shall now be used in an attempt to obtain a better background model following the procedure outlined above.

\subsubsection{First scenario: Large systematic effects}

\begin{figure}[tbp]
  \begin{center}
    \includegraphics[width=.7\textwidth]{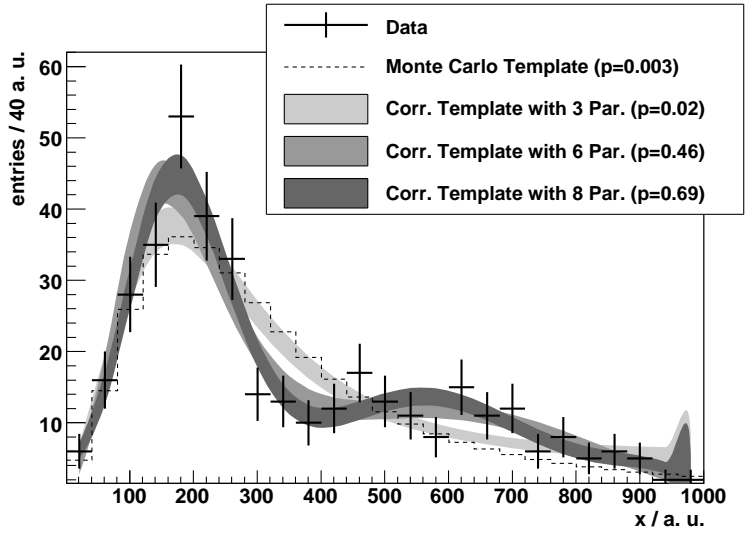}
  \end{center}
  \caption{The Monte Carlo estimate for the data distribution on the left of figure \protect\ref{fig:IntroPlot} is modified with correction functions of
an increasing number of parameters until a satisfactory goodness-of-fit is reached, expressed by the $p$-value of equation (\protect\ref{eq:pval}).}
\label{fig:ModelSelectionPlot}
\end{figure}

\paragraph{Choosing the best polynomial.}
For this example ordinary polynomials are taken as correction functions. Starting with the first scenario, the central prediction ("zeroth-order model") is modified with polynomials of order 2, 5, and 7, displayed in increasing shades of grey in figure \ref{fig:ModelSelectionPlot}. The width of the bands 
corresponds to the respective statistical uncertainty. Table \ref{tab:ModelSelectionTable} shows both types of $p$-values for functions up to order 10. Using a correction function of degree 5 yields the first model with an acceptable goodness-of-fit of 0.46, which can be 
improved even further by including more parameters.

The compatibility peaks at a value of 0.69 when using 8 adjustable parameters. The decline for more complex models results from increasing the number of
free parameters while not gaining a substantial improvement in terms of data description, as is also reflected by a high $p(q_{8,9})$ value of about 0.8. As was 
proposed above the model with the highest $p$-value is taken as the new improved background model.

\begin{table}[tbp]
\begin{center}
	\caption{ $p$-values for the test statistics defined in equations (\protect\ref{eq:qabs}) and (\protect\ref{eq:qrel}) for the data distribution shown in figure  \protect\ref{fig:ModelSelectionPlot}. 
	     }\label{tab:ModelSelectionTable}
	\vspace{5pt}
	\begin{tabular}{|l|c|c|} \hline
	Type of correction function & $p(q_{\vec{\nu}})$ & $p(q_{m,m+1})$ \\
	\hline
	none (fixed to unity) & 0.0027 & 0.15 \\ 
	Constant & 0.0033 & 0.17 \\ 
	Linear & 0.0038 & 0.0075 \\ 
	Quadratic & 0.018 & 0.019 \\ 
	Cubic & 0.052 & 0.0013 \\ 
	4th degree & 0.33 & 0.072 \\
	5th degree & 0.46 & 0.34 \\
	6th degree & 0.46 & 0.04 \\
	7th degree & 0.69 & 0.80 \\
	8th degree & 0.63 & 0.21 \\
	9th degree & 0.68 & 0.99 \\
	10th degree & 0.60 & - \\
	\hline
	\end{tabular}
\end{center}
\end{table}

\paragraph{Alternative starting templates.}

\begin{figure}[tbp]
\begin{center}

\subfigure{
\includegraphics[width= 
0.45\textwidth]{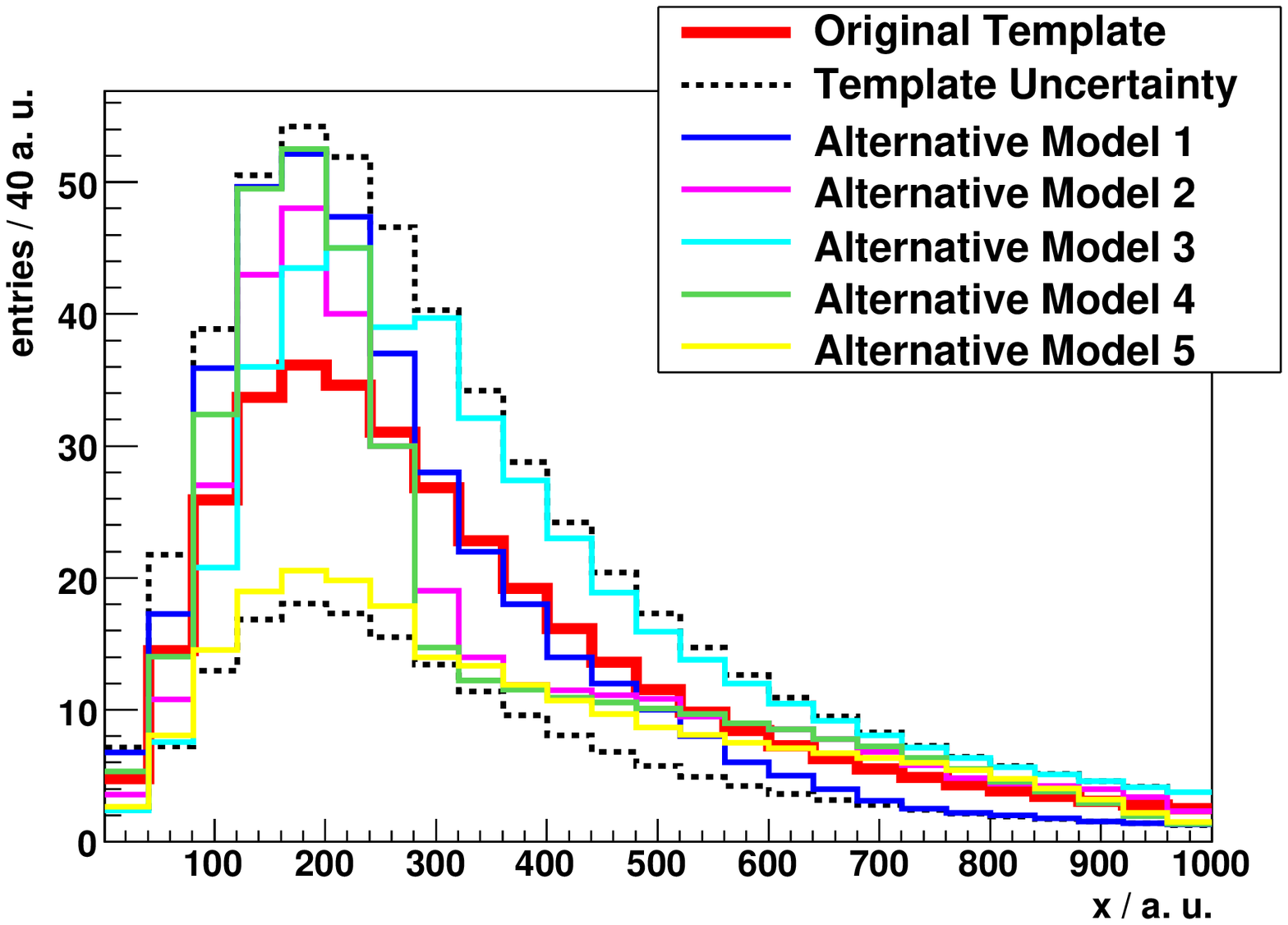}
}
\subfigure{
\includegraphics[width= 
0.45\textwidth]{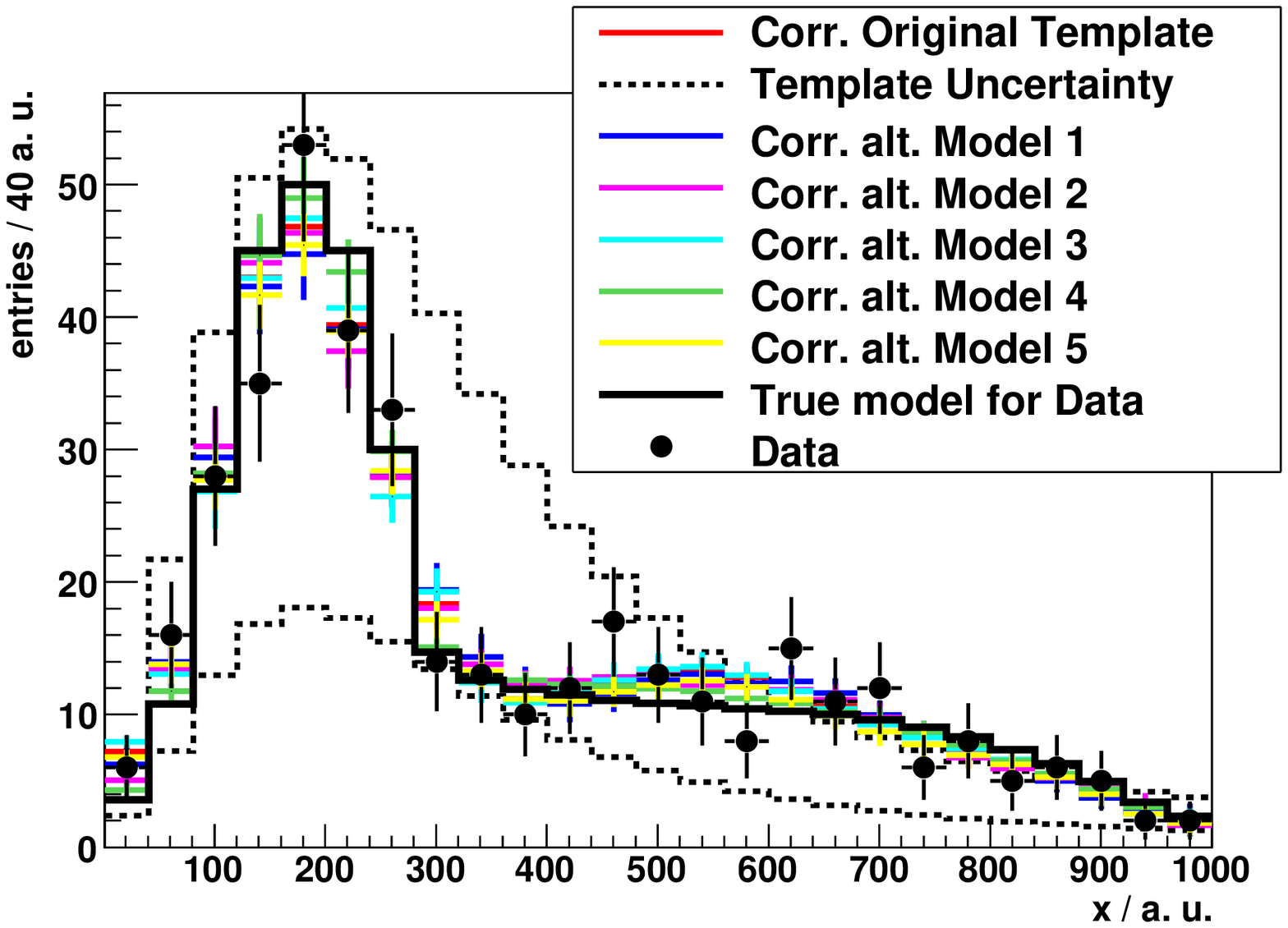}
}
\caption{ Selecting different templates within the systematic uncertainty of the original Monte Carlo prediction to estimate their influence
 on the corrected model.
}\label{fig:VariousTemplates}
\end{center}
\end{figure}

\begin{figure}[tbp]
\begin{center}

\subfigure{
\includegraphics[width= 
0.45\textwidth]{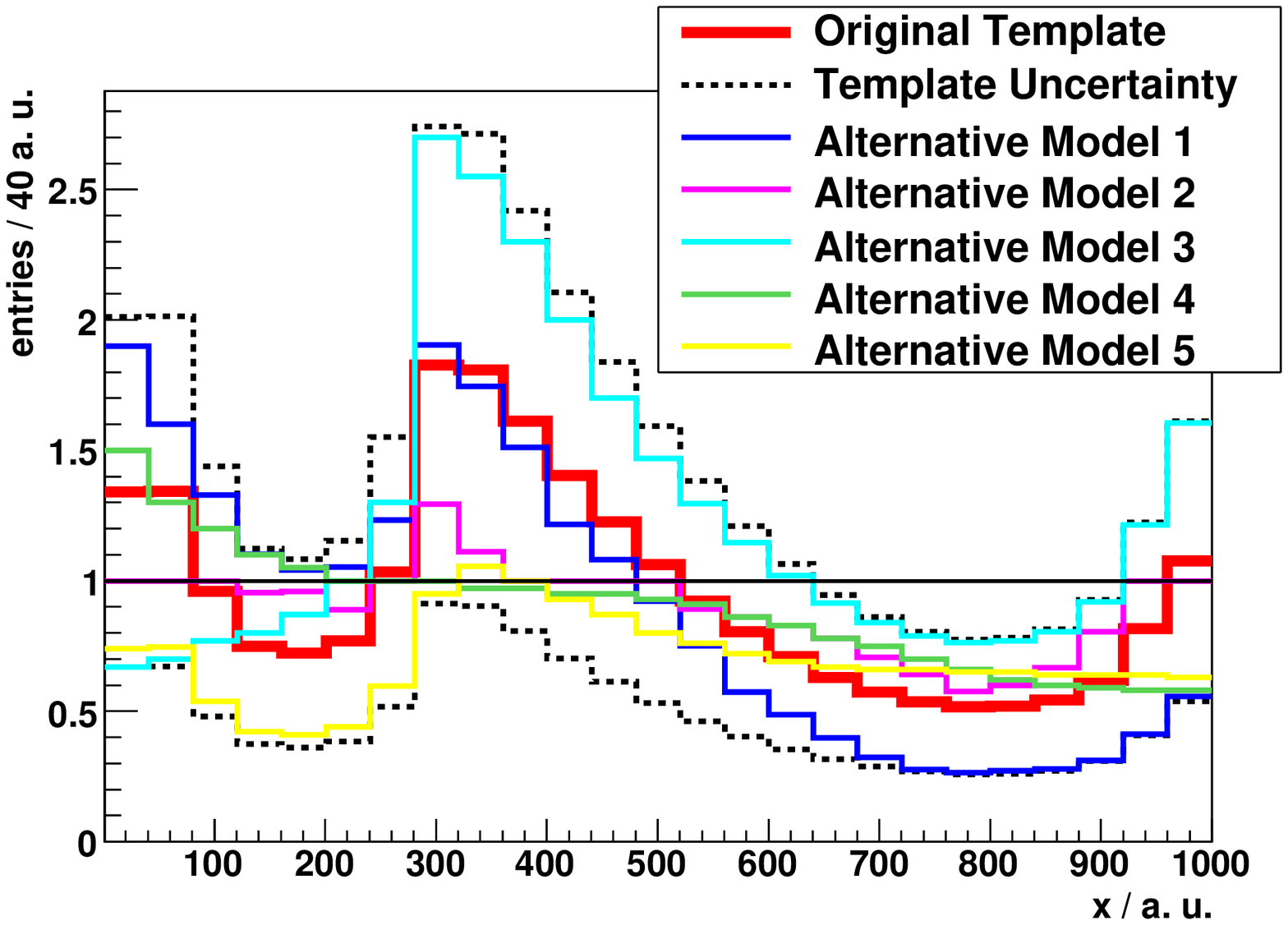}
}
\subfigure{
\includegraphics[width= 
0.45\textwidth]{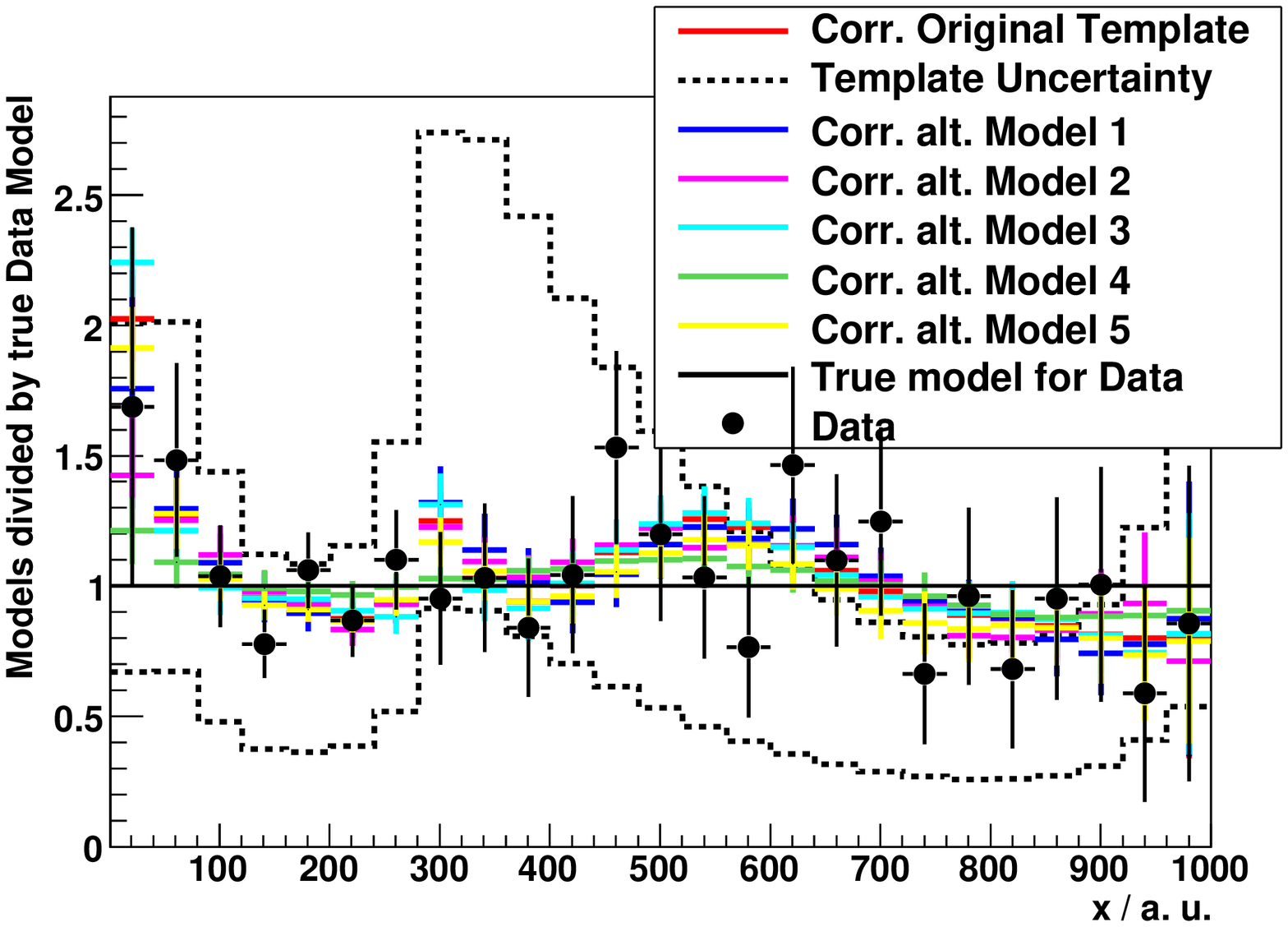}
}
\caption{ Models before (left) and after (right) correction divided by the true model for data.
}\label{fig:RatioVariousTemplates}
\end{center}
\end{figure}

Apart from the statistical error of the fit, an additional uncertainty arises from the choice of the starting template. To investigate the dependency of the corrected model on a particular shape of the original hypothesis, additional starting templates are selected from within
the systematic uncertainty of the MC prediction, as shown in the left plot of figure \ref{fig:VariousTemplates}. As was mentioned above, in the case of real data one would vary the MC prediction according to known systematic effects, thereby obtaining a set of possible starting templates. All those templates are 
corrected separately with the polynomial yielding the highest absolute goodness-of-fit, shown in the right plot of the figure. In addition, the true model from which the data
were generated is displayed as the black solid line. After correction the new models nicely converge to the true model, almost regardless of the shape of the starting template. Figure \ref{fig:RatioVariousTemplates} shows the same curves divided by the true data model. The strong deviations of the original prediction (thick red line) from the constant line at unity represent the rather extreme introduced systematic effects, which call for a correction using 
several parameters.

\paragraph{New background model and its uncertainty.}
Assuming a flat prior probability for the different starting templates, the best estimated model is finally taken as the mean value of all corrected templates. 
Its total uncertainty is calculated by generating 2000 toy data sets from this estimate and applying the proposed method on every one of them.

\begin{figure}[tbp]
  \begin{center}
    \includegraphics[width=.7\textwidth]{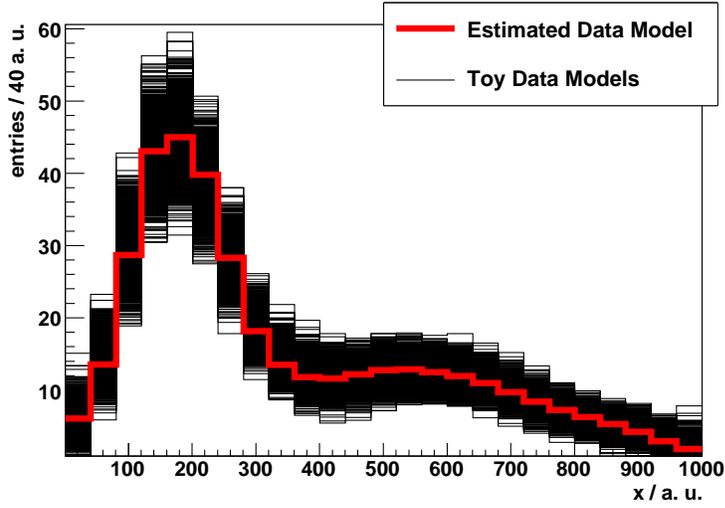}
  \end{center}
  \caption{ Corrected models using 2000 sets of toy data generated from the "best estimated model" shown in red. The RMS of each bin is taken as an 
estimate for the statistical uncertainty of the method.
  }\label{fig:AllToyModels}
\end{figure}

\begin{figure}[tbp]
  \begin{center}
    \includegraphics[width=.7\textwidth]{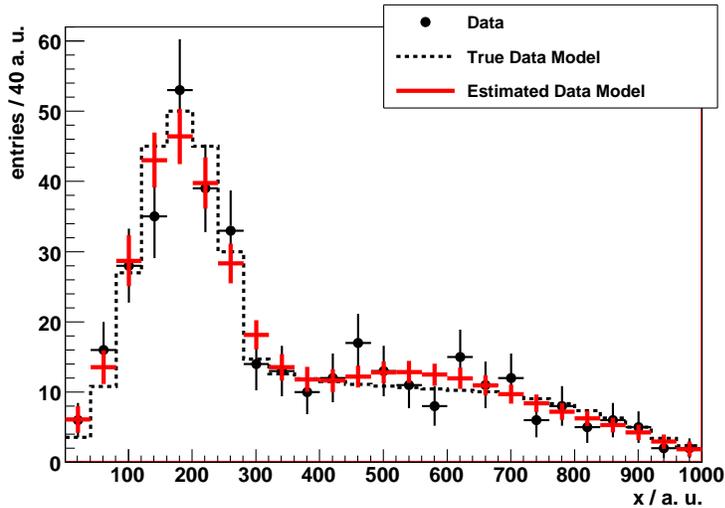}
  \end{center}
  \caption{ The estimated and the true model for the data agree well within the indicated uncertainty.
  }\label{fig:EstimatedAndTrueModel}
\end{figure}

The bin-wise RMS of the corrected models' distribution (see figure \ref{fig:AllToyModels}) together with the inter-bin correlation is then taken as an estimate for the statistical error. Figure \ref{fig:EstimatedAndTrueModel} contrasts the best estimated model with the true model for the data. The true model is nicely reproduced.

\subsubsection{Second scenario: No systematic effects}

\begin{figure}[htbp]
  \begin{center}
    \includegraphics[width=.7\textwidth]{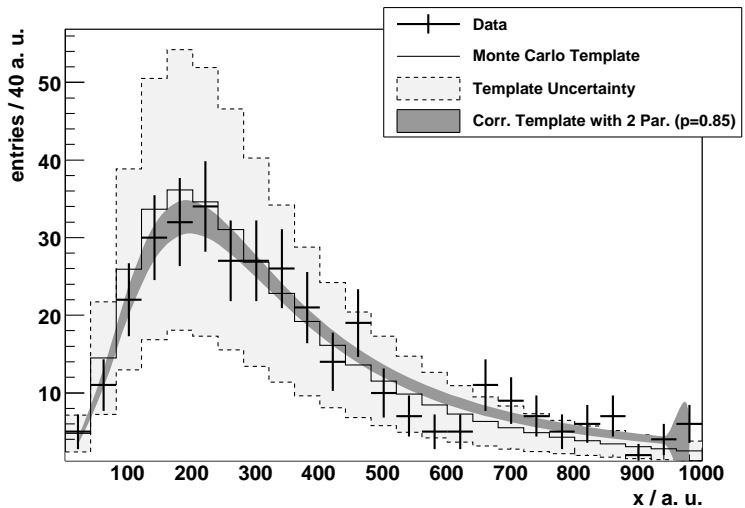}
  \end{center}
  \caption{ The Monte Carlo estimate for the data distribution on the right of figure \protect\ref{fig:IntroPlot} is modified with correction functions of
an increasing number of parameters until a satisfactory goodness-of-fit is reached, expressed by the $p$-value of equation (\protect\ref{eq:pval}). In this scenario
two parameters are sufficient yielding a $p$-value of about 85\%.
  }\label{fig:ModelSelectionPlotNoSys}
\end{figure}

The second scenario shown on the right of figure \ref{fig:TwoScenarios} shall illustrate the usefulness of the proposed method when there is apparently only 
little or no systematic deviation present. As before, first the central MC prediction is modified with the correction function giving the highest $p(q_{\vec{\nu}})$
value. This turns out to be a linear function (see table \ref{tab:ModelSelectionTable2}) which slightly tilts the template as can be seen in figure  \ref{fig:ModelSelectionPlotNoSys}. It is noteworthy that a simple rescaling does not improve the data description ($p(q_{0,1}) = 0.96$) in this case, whereas the linear correction returns a significantly better model, as demonstrated by  $p(q_{1,2}) = 0.01$.

\begin{table}[htbp]
\begin{center}
	\caption{ $p$-values for the test statistics defined in equations (\protect\ref{eq:qabs}) and (\protect\ref{eq:qrel}) for the data distribution shown in figure  \protect\ref{fig:ModelSelectionPlotNoSys}. The templates are multiplied
	with polynomials of the degree stated in the first row.}
	\label{tab:ModelSelectionTable2}
	\vspace{5pt}
	\begin{tabular}{|l|c|c|} \hline
	Type of correction function & $p(q_{\vec{\nu}})$ & $p(q_{m,m+1})$ \\
	\hline
	none (fixed to unity) & 0.56 & 0.96 \\ 
	Constant & 0.51 & 0.01 \\ 
	Linear & 0.85 & 0.41 \\ 
	Quadratic & 0.84 & 0.40 \\ 
	Cubic & 0.83 & 0.90 \\ 
	4th degree & 0.79 & 0.38 \\
	5th degree & 0.78 & 0.46 \\
	6th degree & 0.76 & 0.13 \\
	7th degree & 0.84 & 0.37 \\
	8th degree & 0.80 & 0.79 \\
	9th degree & 0.74 & 0.36 \\
	10th degree & 0.67 & - \\
	\hline
	\end{tabular}
\end{center}
\end{table}

\begin{figure}[htbp]
\begin{center}

\subfigure{
\includegraphics[width= 
0.45\textwidth]{./figures/VarTemplSys.eps}
}
\subfigure{
\includegraphics[width= 
0.45\textwidth]{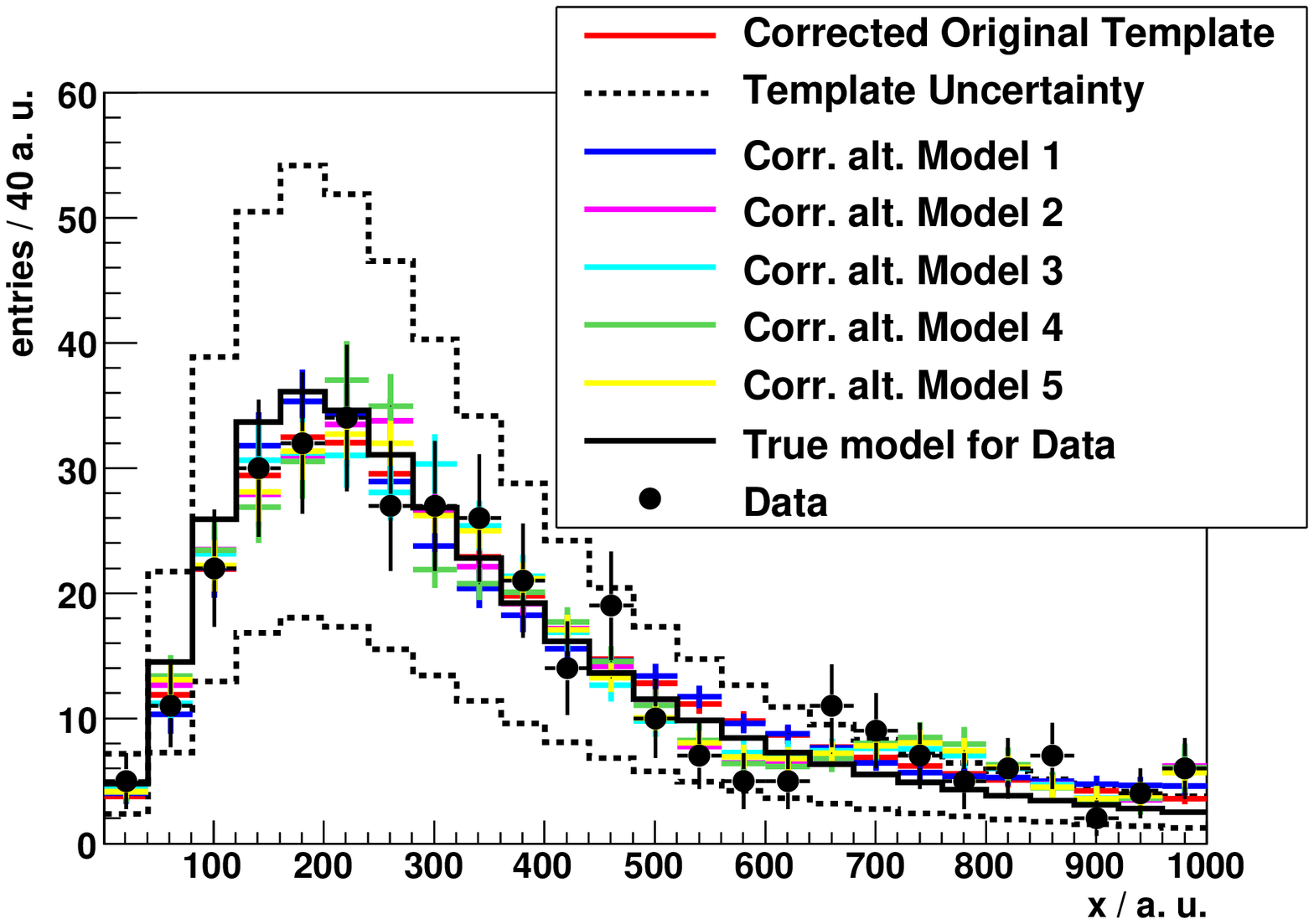}
}
\caption{ Selecting different templates within the systematic uncertainty of the original Monte Carlo prediction to estimate their influence
 on the corrected model.
}\label{fig:VariousTemplatesNoSys}
\end{center}
\end{figure}

\begin{figure}[htbp]
\begin{center}

\subfigure{
\includegraphics[width= 
0.45\textwidth]{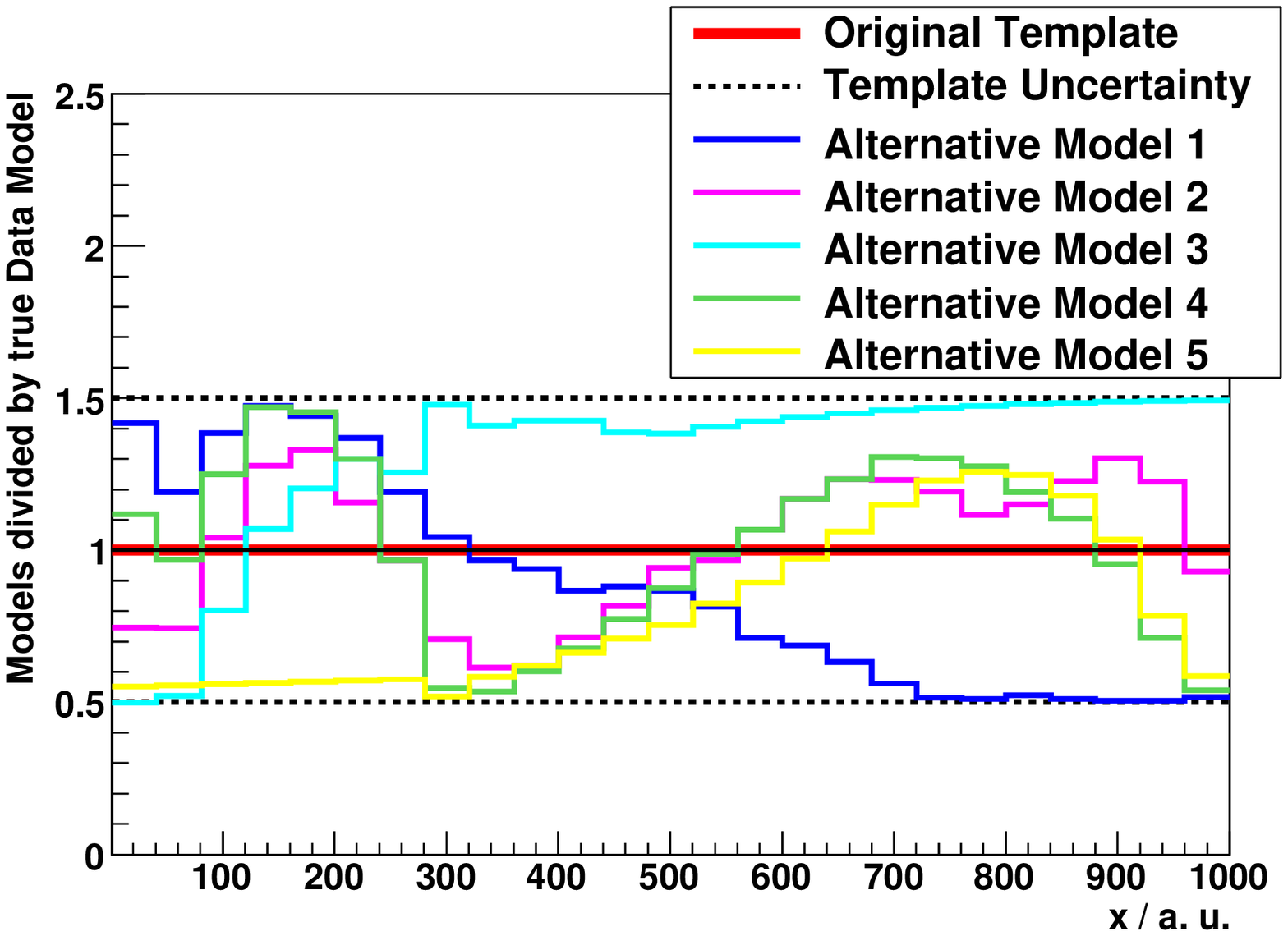}
}
\subfigure{
\includegraphics[width= 
0.45\textwidth]{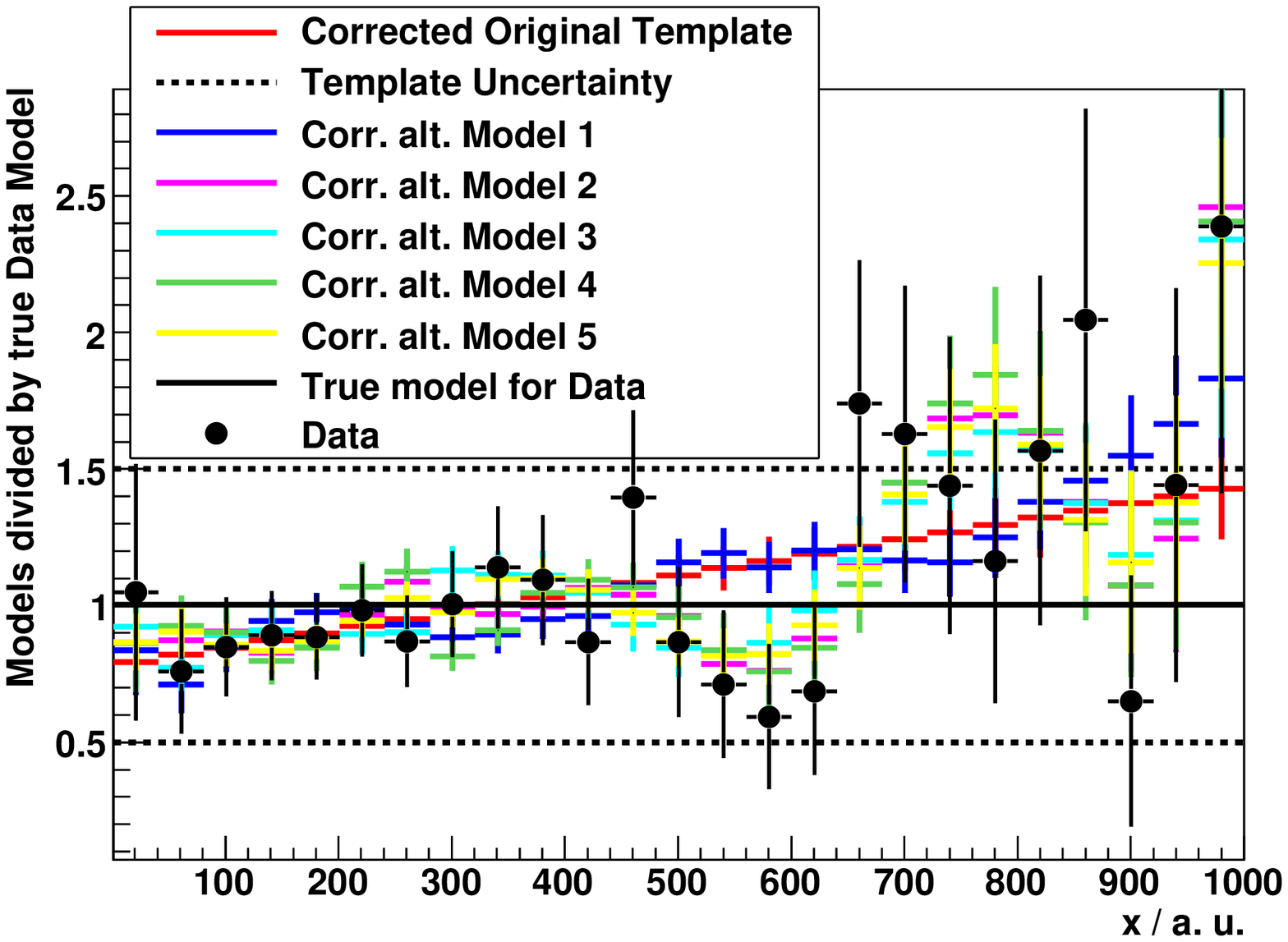}
}
\caption{ Models before (left) and after (right) correction divided by the true model for data.
}\label{fig:RatioVariousTemplatesNoSys}
\end{center}
\end{figure}

\begin{figure}[htbp]
  \begin{center}
    \includegraphics[width=.7\textwidth]{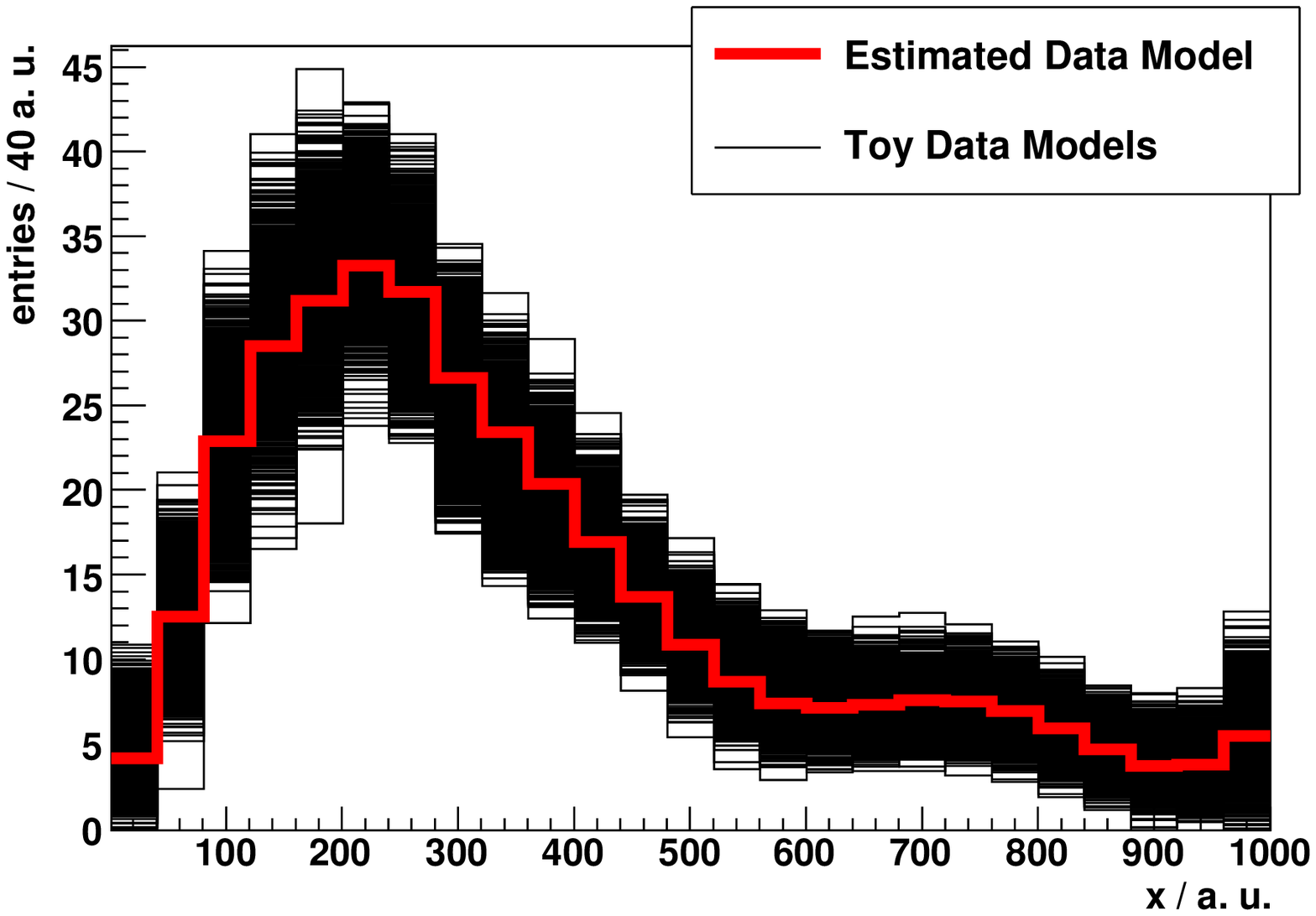}
  \end{center}
  \caption{ Corrected models using 2000 sets of toy data generated from the "best estimated model" shown in red. The RMS of each bin is taken as an 
estimate for the statistical uncertainty of the method.
  }\label{fig:AllToyModelsNOSys}
\end{figure}

\begin{figure}[htbp]
  \begin{center}
    \includegraphics[width=.7\textwidth]{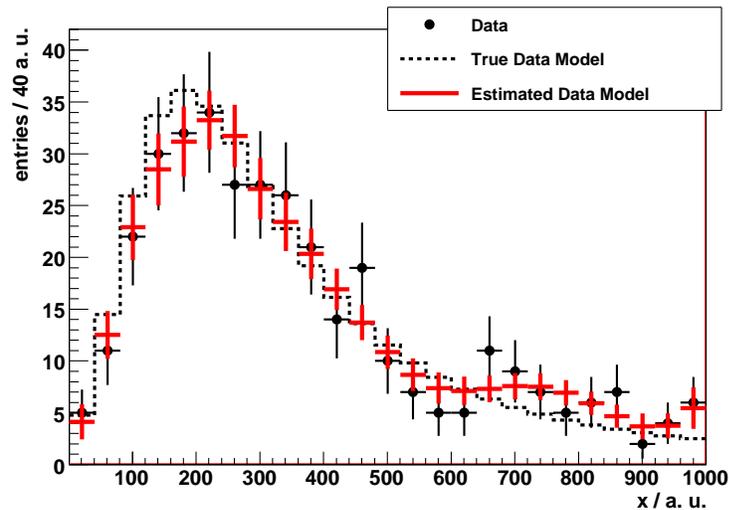}
  \end{center}
  \caption{ The estimated and the true model for the data agree well within the indicated uncertainty.
  }\label{fig:EstimatedAndTrueModelNOSys}
\end{figure}

Figures \ref{fig:VariousTemplatesNoSys} and \ref{fig:RatioVariousTemplatesNoSys} show again the correction of different starting templates with respect to the true model for data. As a limiting case, this scenario was generated without systematic effects. Hence the central prediction would constitute the best model.
Still, the proposed method has the various templates converge to the true model. The offset at higher x-values results from a bias introduced by the data. Again, toy experiments are generated from the mean of the corrected models to obtain the statistical uncertainty (see figure \ref{fig:AllToyModelsNOSys}). Figure \ref{fig:EstimatedAndTrueModelNOSys} 
shows the estimated and the true model for the data which agree well within the indicated uncertainty. 

\subsubsection{Extrapolation to signal region}

Once the improved background model has been determined following the procedure described above
the same correction is to be applied on the Monte Carlo prediction for the background in the 
signal region (see also section 4). In addition, systematic effects associated with the transfer of the correction 
from control to signal region need to be considered, such as the different influence on the shapes
of the distributions by certain systematic sources. These have to be treated on a case-by-case 
basis and lie beyond the scope of this paper. 

After establishing a signal one would want to quote
its cross-section. It is very likely that the same correction for systematic effects would also
make a reliable signal Monte Carlo prediction more realistic. The proposed method also considerably facilitates the search for a suitable control region, since 
differences between signal and control region due to uncontroversial features of the simulation
(phase space, for instance) are accounted for automatically.

After a digression introducing another useful basis set, section 4 demonstrates the performance of the
proposed method with respect to two other common background estimation techniques.

\subsection{Bernstein polynomials}

This subsection discusses the use of a different set of correction functions, 
that of Bernstein polynomials. In principle, using Bernstein polynomials is
equivalent to using ordinary polynomials. Nevertheless, due to their more localized
effect on the original model (see below) their usage may result in a smaller 
correlation between the fitted parameters.

\begin{figure}[htbp]
\begin{center}

\subfigure{
\includegraphics[width= 
0.30\textwidth]{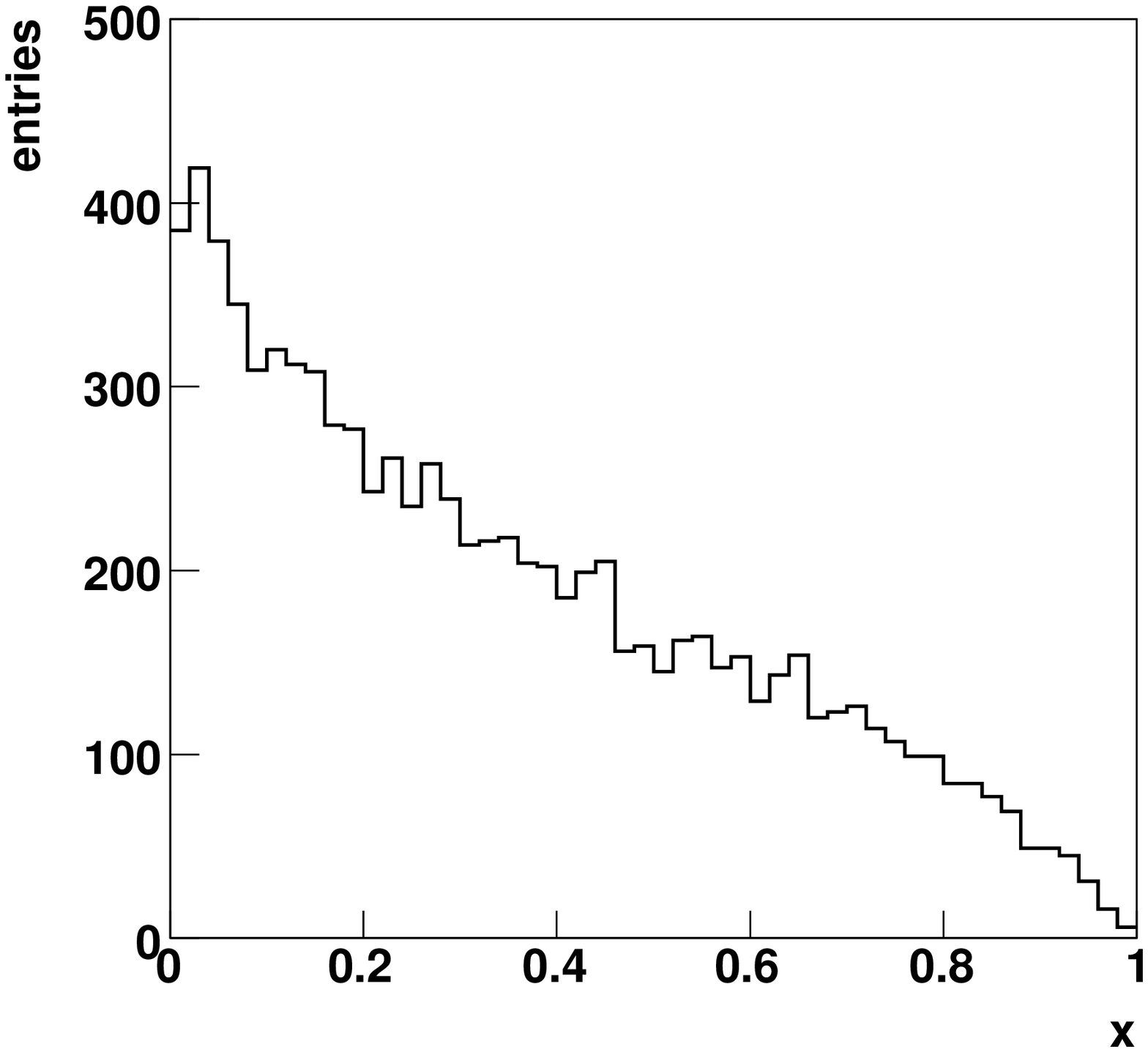}
}
\subfigure{
\includegraphics[width= 
0.30\textwidth]{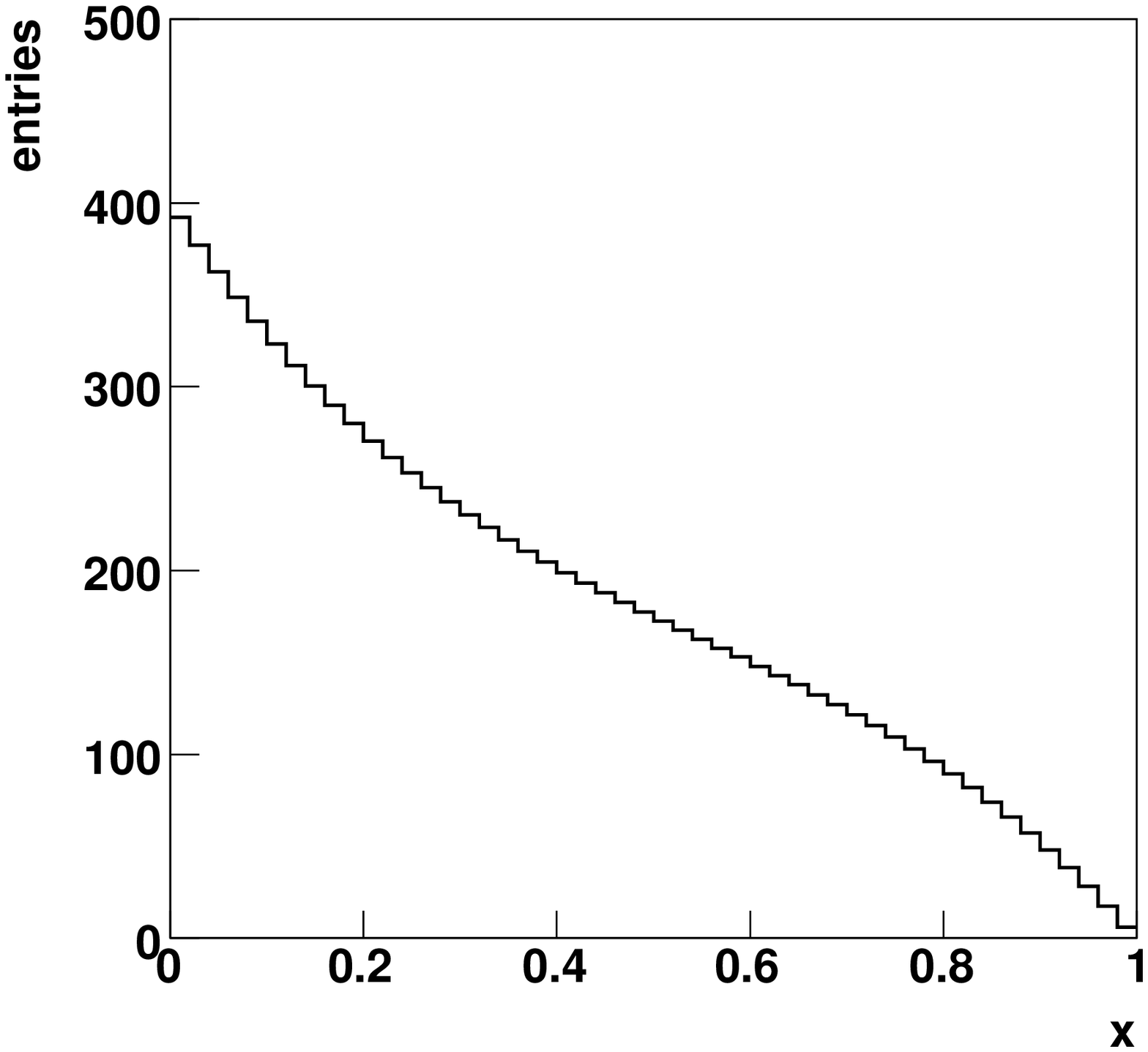}
}
\subfigure{
\includegraphics[width= 
0.30\textwidth]{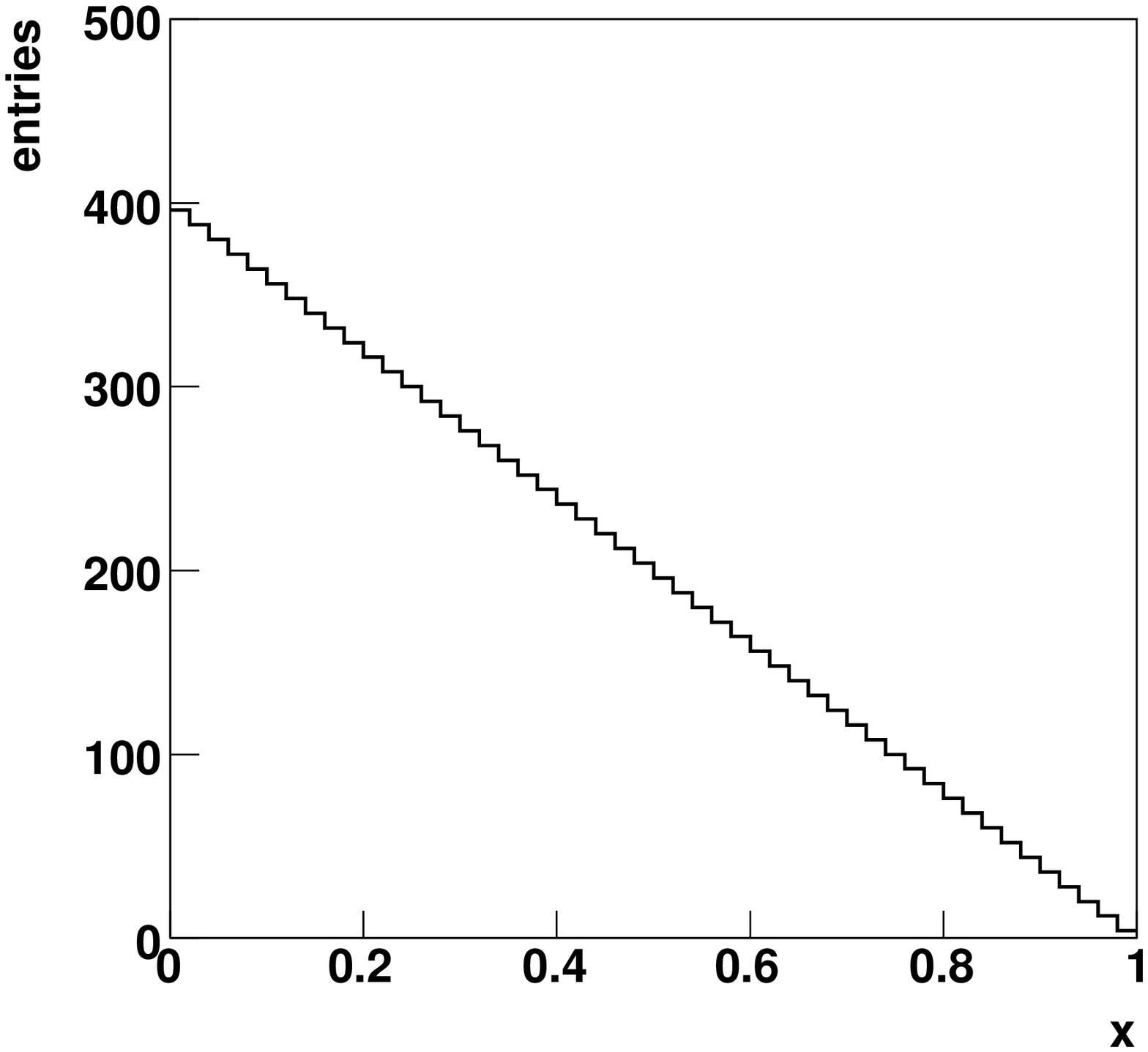}
}
\caption{ Data (left plot) originating from a true model (middle) and the 
corresponding estimate from a Monte Carlo simulation (right).
}\label{fig:dist}
\end{center}
\end{figure}

Suppose that data, which is shown in the left plot of figure \ref{fig:dist},
originate from the true model displayed in the middle plot, whereas the Monte Carlo
estimate predicts a simple slope shown at the right.

\begin{figure}[htbp]
\begin{center}

\subfigure{
\includegraphics[width= 
0.30\textwidth]{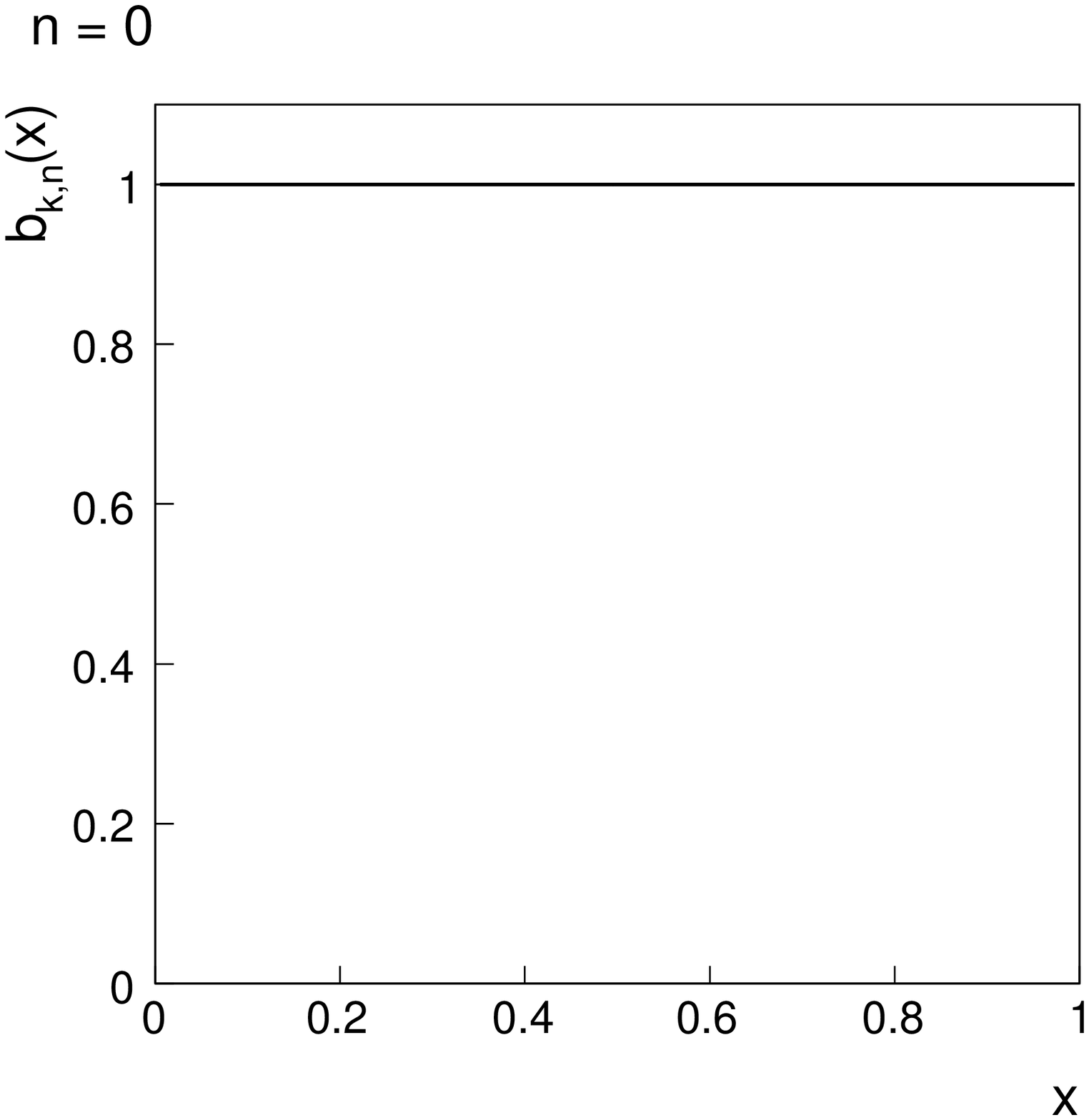}
}
\subfigure{
\includegraphics[width= 
0.30\textwidth]{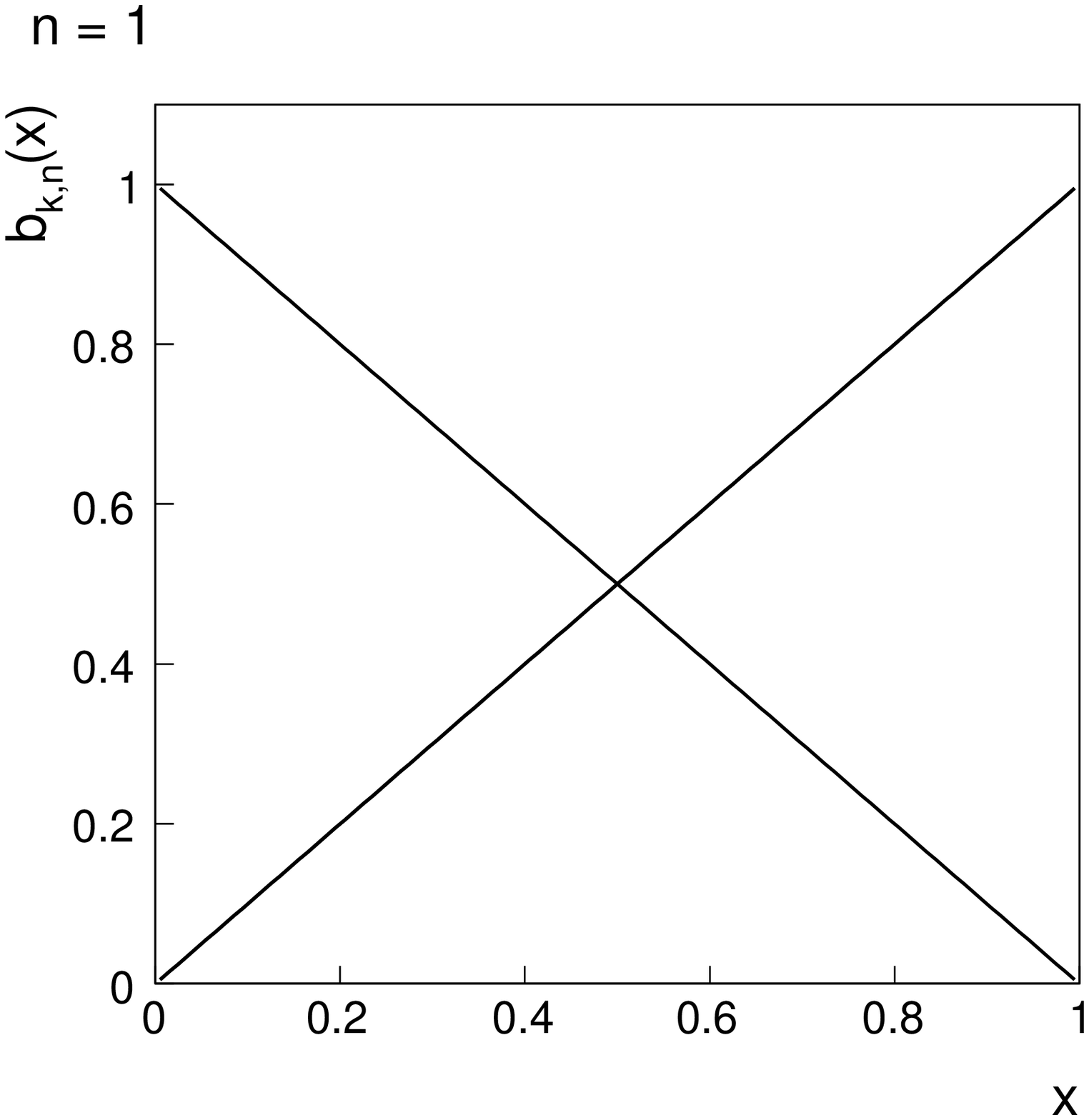}
}
\subfigure{
\includegraphics[width= 
0.30\textwidth]{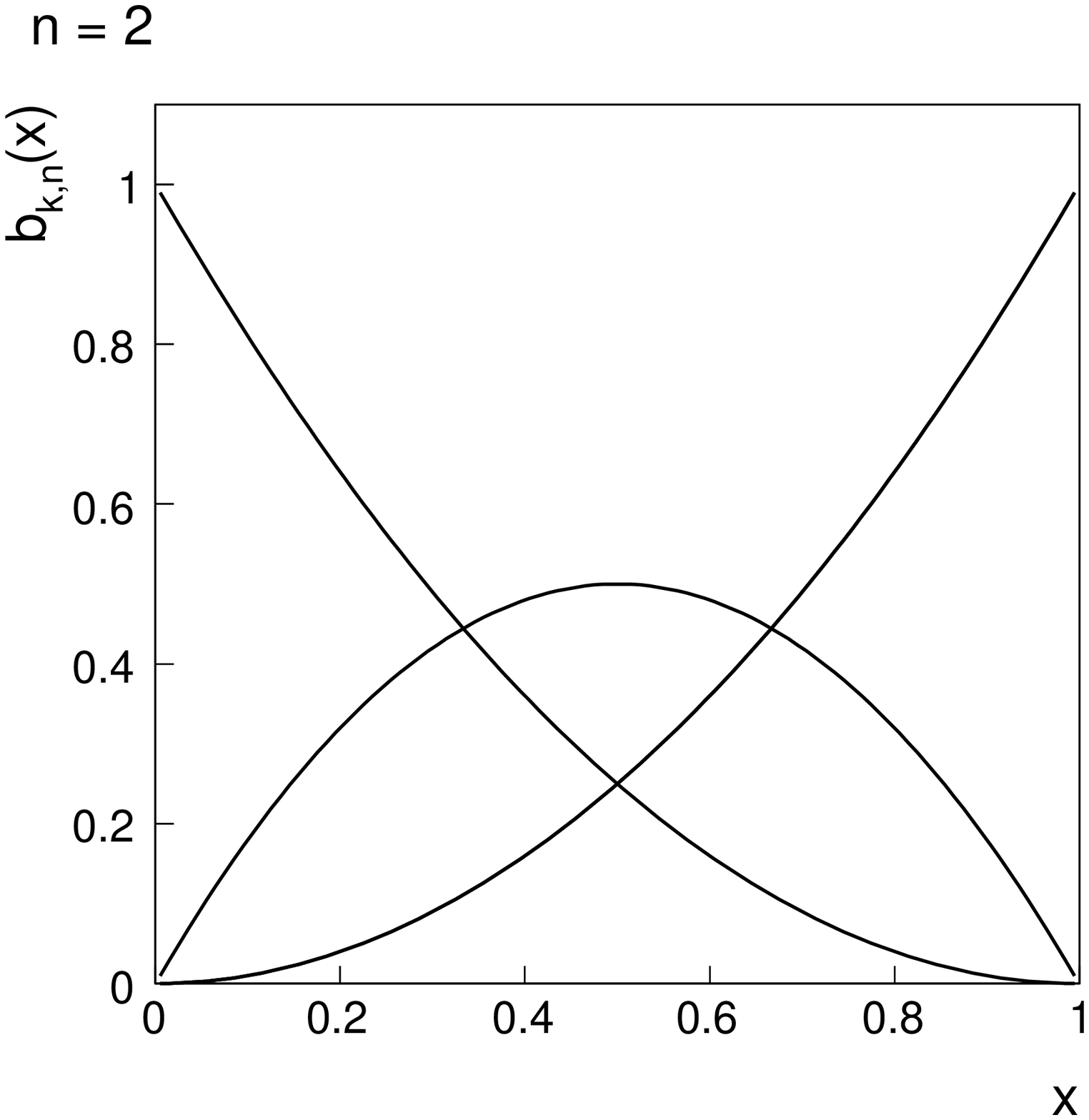}
}
\subfigure{
\includegraphics[width= 
0.30\textwidth]{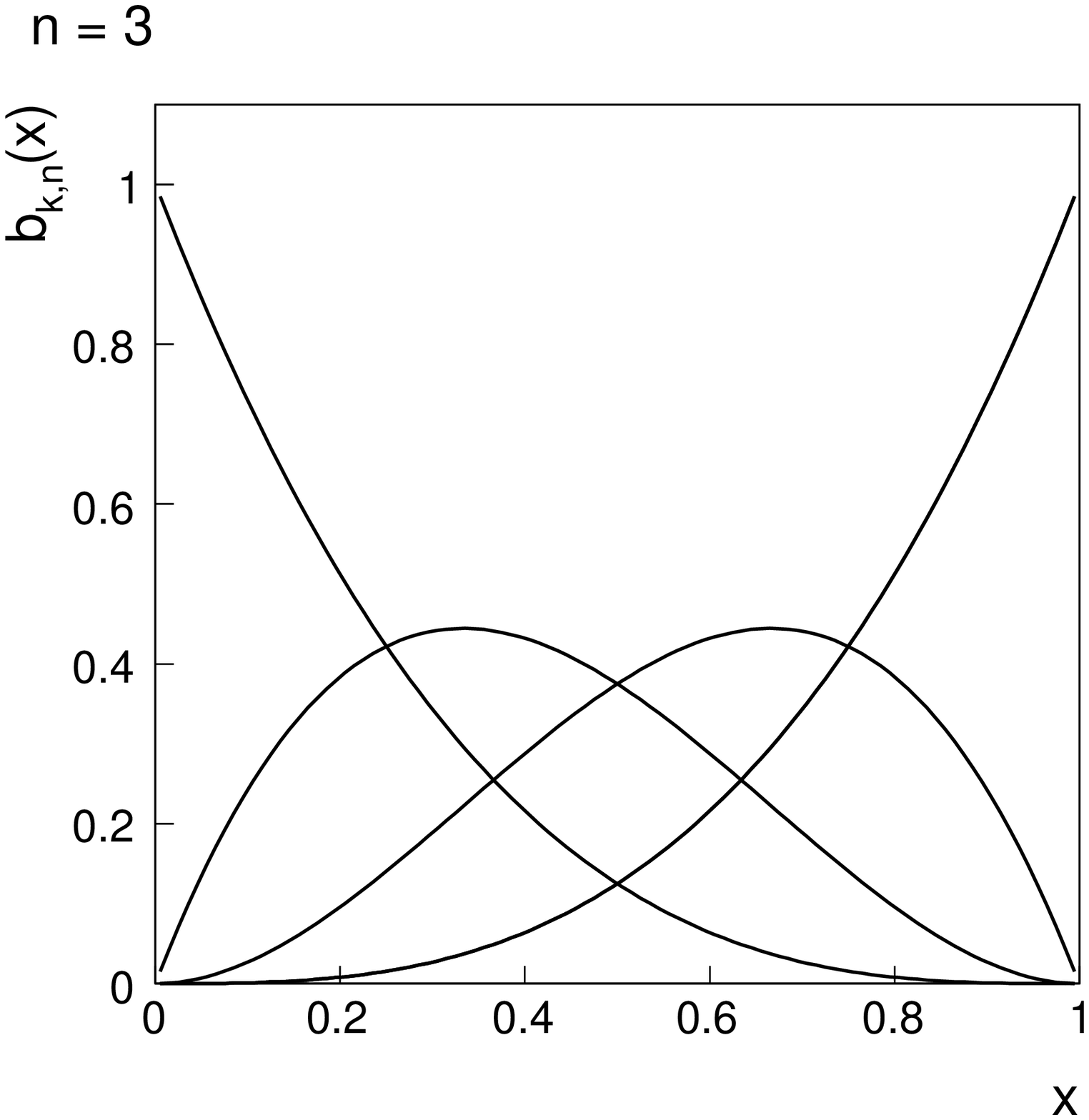}
}
\subfigure{
\includegraphics[width= 
0.30\textwidth]{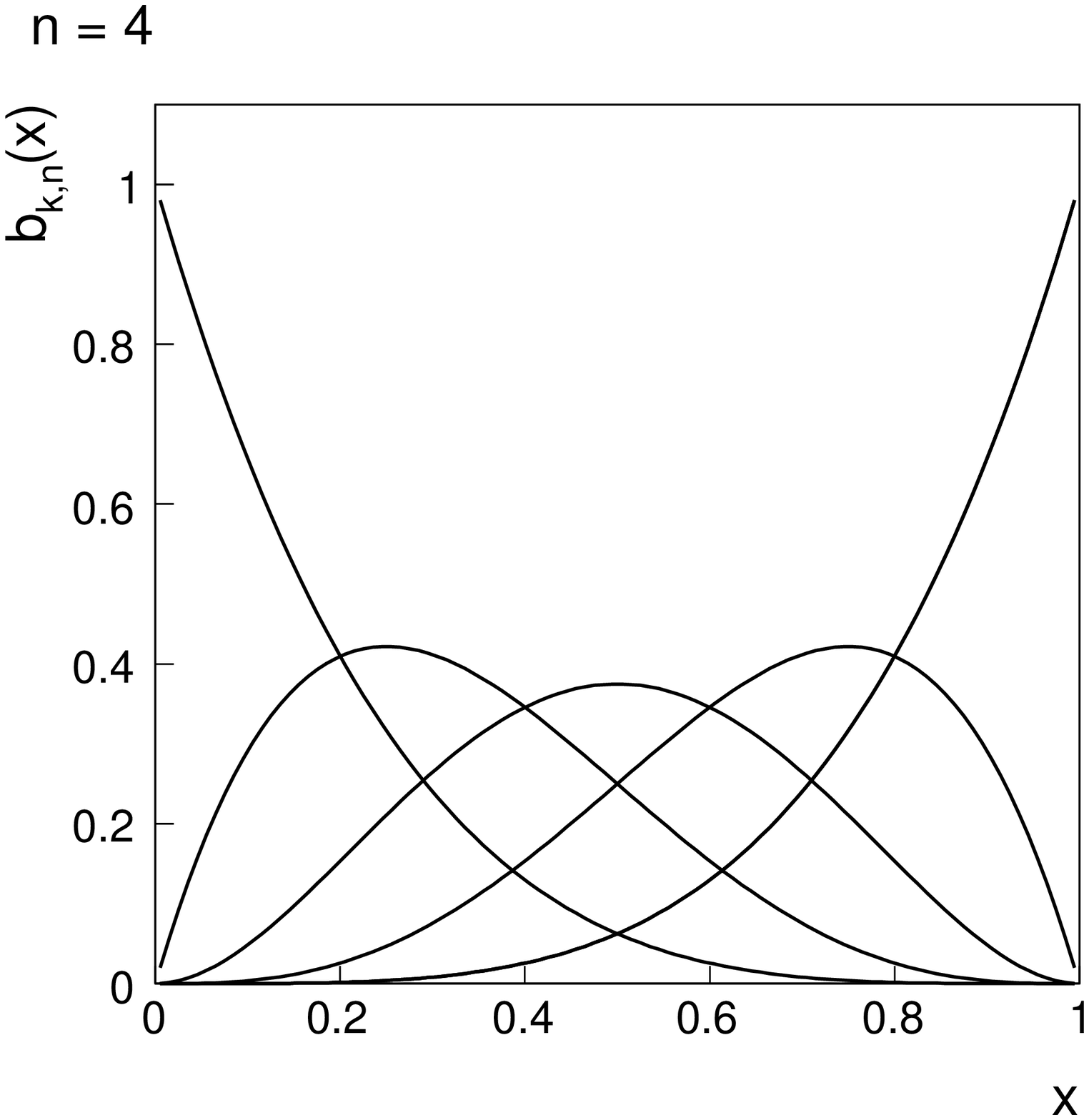}
}
\subfigure{
\includegraphics[width= 
0.30\textwidth]{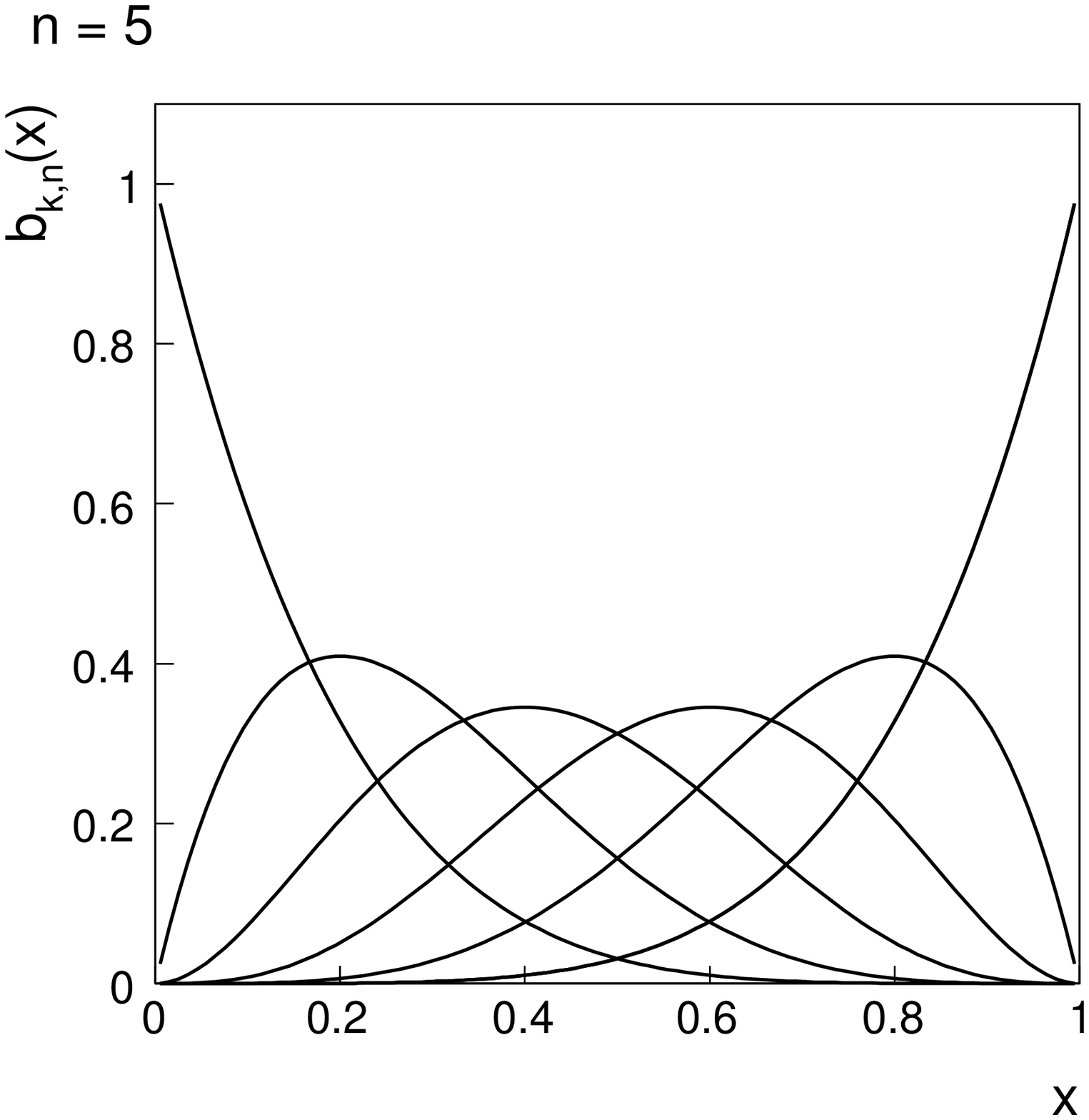}
}

\caption{ Bernstein basis polynomials of different orders $n$.
\label{fig:bernstein}}
\end{center}
\end{figure}

The Bernstein polynomials are built via a superposition of certain basis 
polynomials (see e.g.~\cite{Bernstein}). The set of $m+1$ Bernstein basis
polynomials of order $m$ are defined as

\begin{equation}
\label{eq:bernsteindef}
b_{k,m}(x) = \frac{m!}{k! (m-k)!} x^k (1-x)^{m-k} \;.
\end{equation}
These are nonzero in the range $[0,1]$, which corresponds
to the range of the variable $x$ in the example of figure~\ref{fig:dist}.
Variables exceeding that boundaries can be translated
and scaled to lie in this region.  The Bernstein basis polynomials
for orders 0 through 5 are shown in figure~\ref{fig:bernstein}.

The correction function $s(x)$ is then taken as a linear combination of
the basis polynomials (i.e., $s$ is a Bernstein polynomial),

\begin{equation}
\label{eq:scalefunc}
s(x) = \sum_{k=0}^m \beta_k b_{k,m}(x) \;.
\end{equation}

\noindent  For $\beta_k = 1$, $k = 0, \ldots, m$, one has $s(x) = 1$,
so it is easy to identify the point in parameter space that corresponds
to no modification.  An important property of Bernstein polynomials is that
a basis polynomial of a given order $m-1$ can always be written in terms of 
those of order $m$:

\begin{equation}
\label{eq:nested}
b_{k,m-1}(x) = \frac{m-k}{m} b_{k,m}(x)  + \frac{k+1}{m} b_{k+1,m}(x) \;.
\end{equation} 

\noindent This means that the Bernstein polynomials defined using
basis functions of successively increasing order form a nested
family.  That is, the model of order $m$ contains as a special case
the model of order $m-1$.

\begin{figure}[htbp]
\begin{center}

\subfigure{
\includegraphics[width= 
0.30\textwidth]{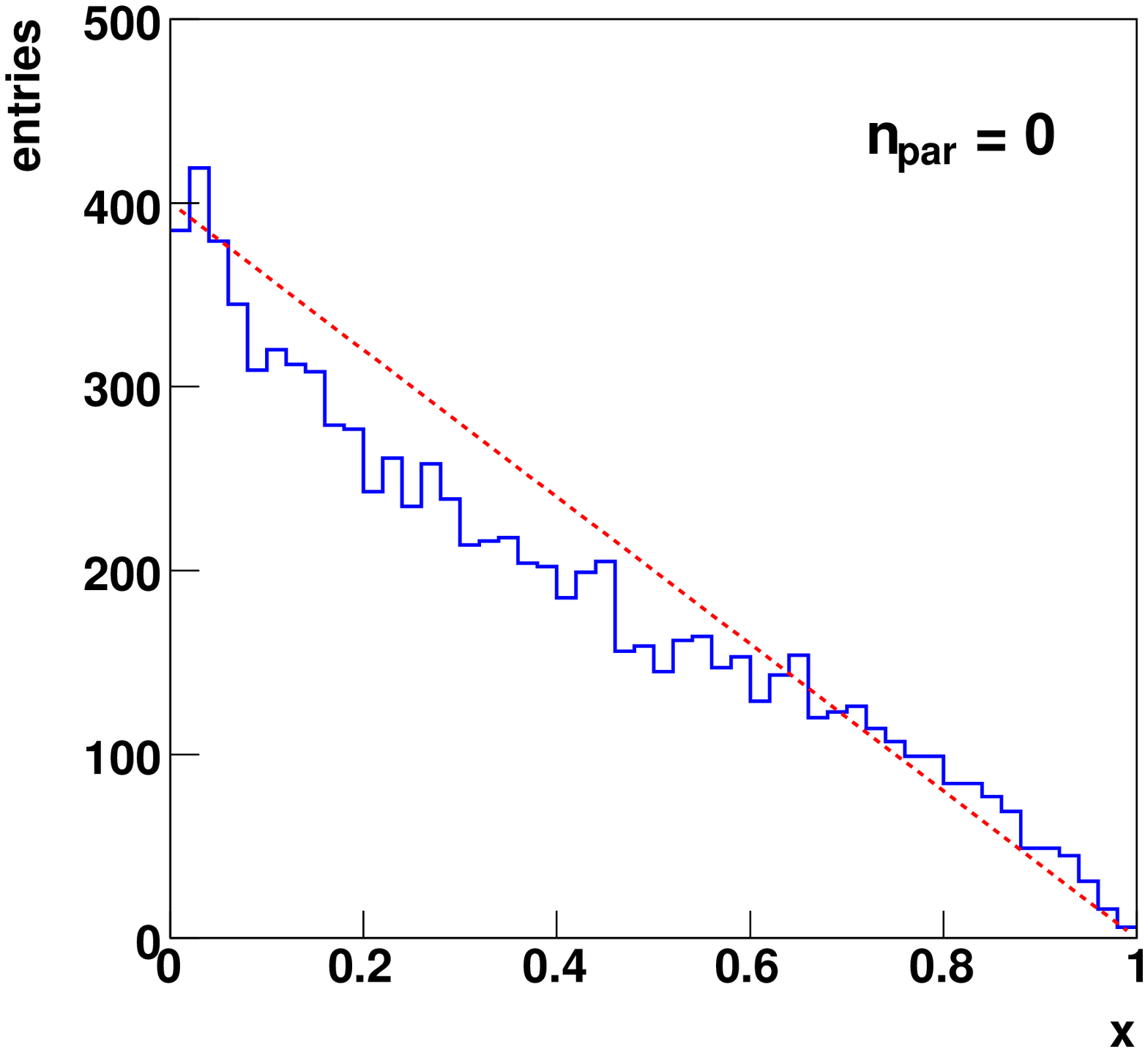}
}
\subfigure{
\includegraphics[width= 
0.30\textwidth]{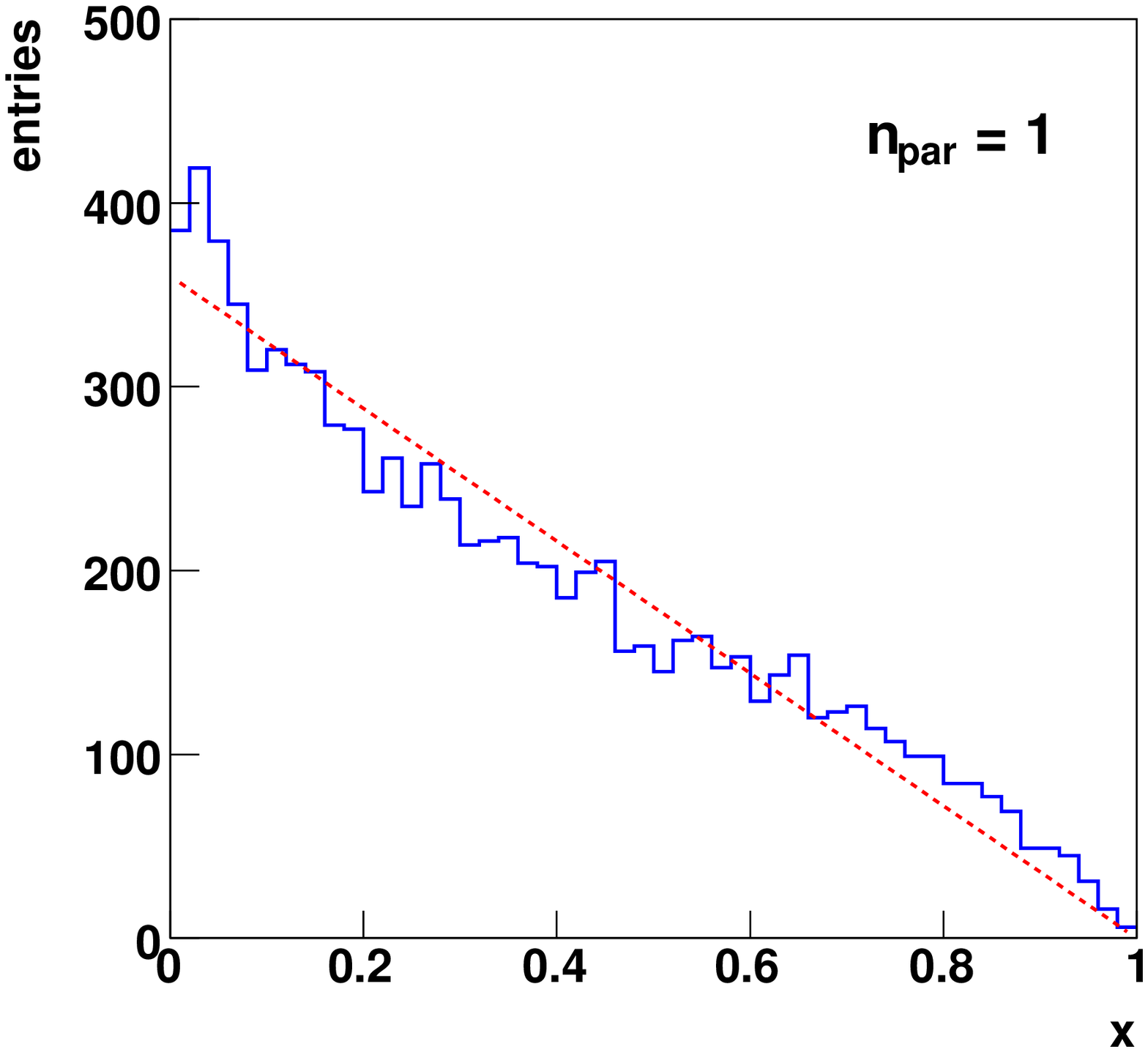}
}
\subfigure{
\includegraphics[width= 
0.30\textwidth]{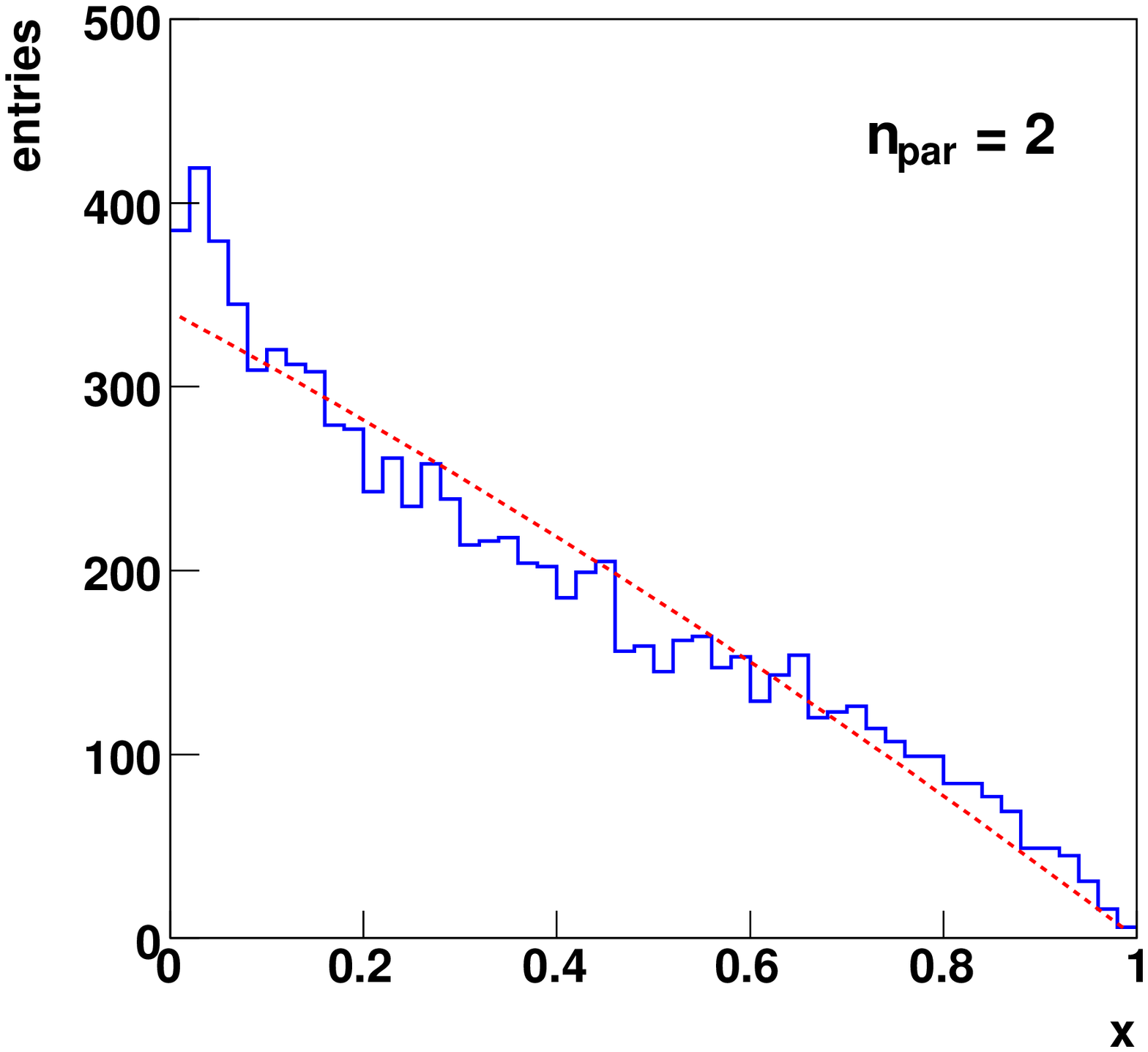}
}
\subfigure{
\includegraphics[width= 
0.30\textwidth]{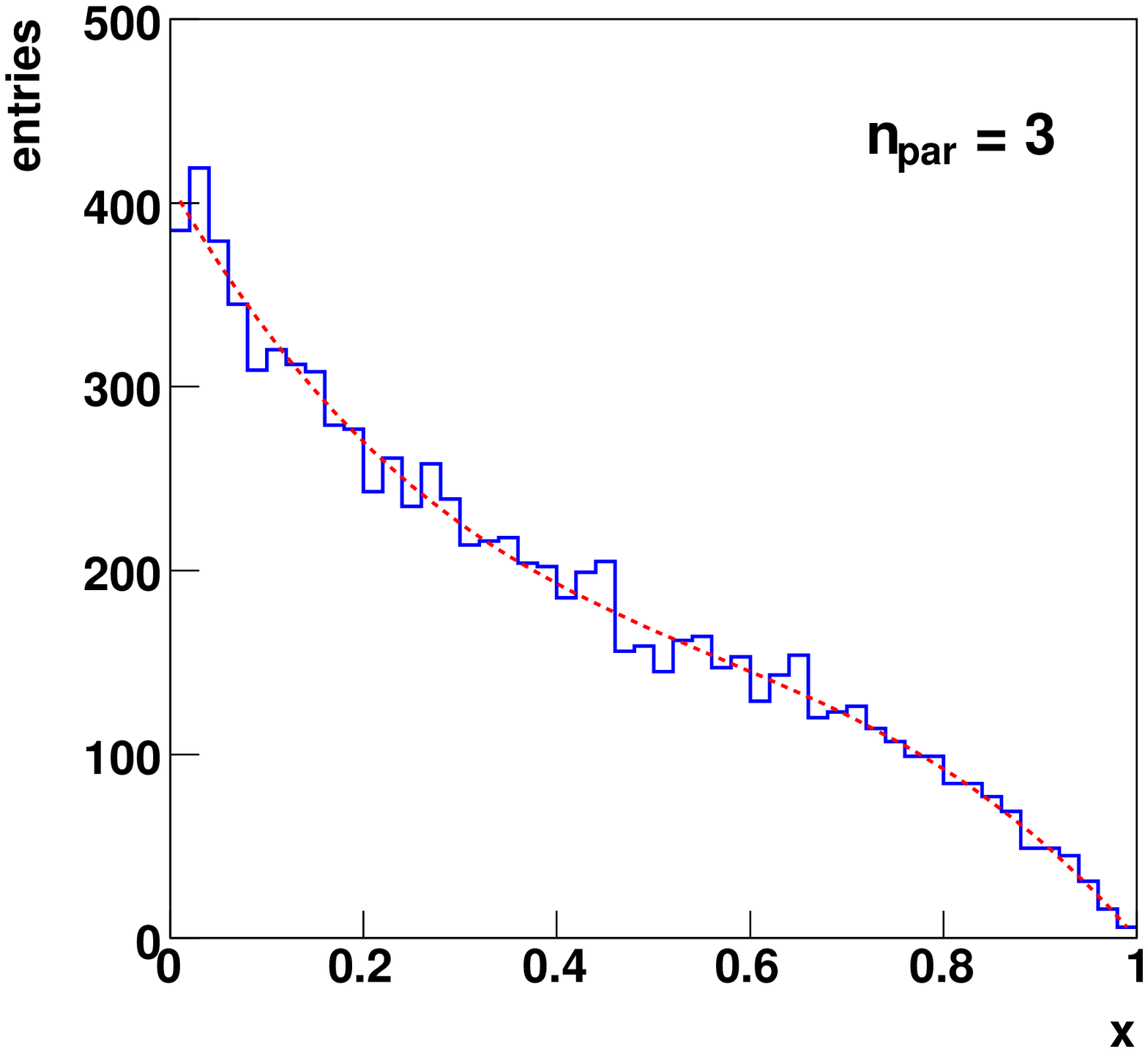}
}
\subfigure{
\includegraphics[width= 
0.30\textwidth]{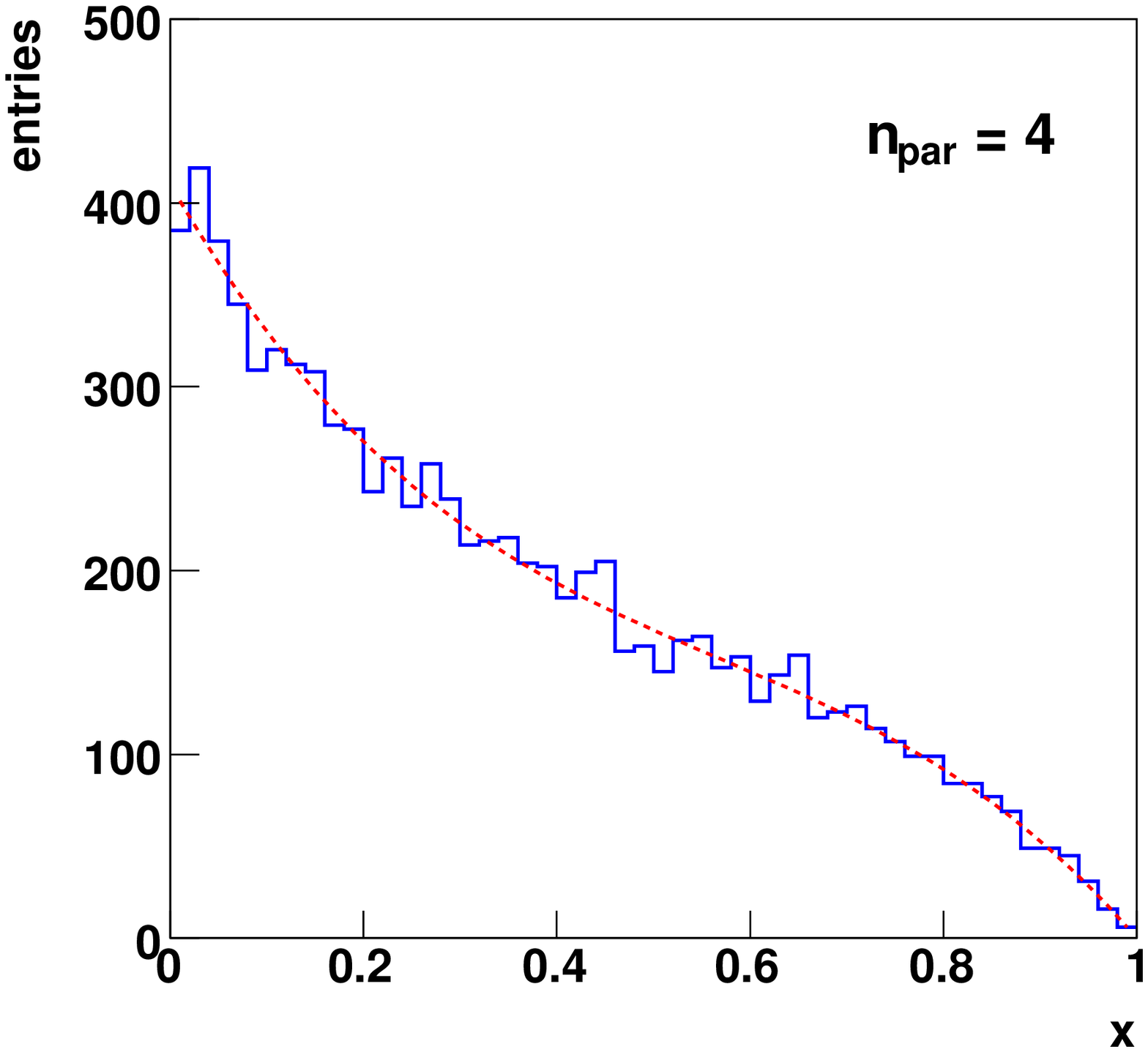}
}
\subfigure{
\includegraphics[width= 
0.30\textwidth]{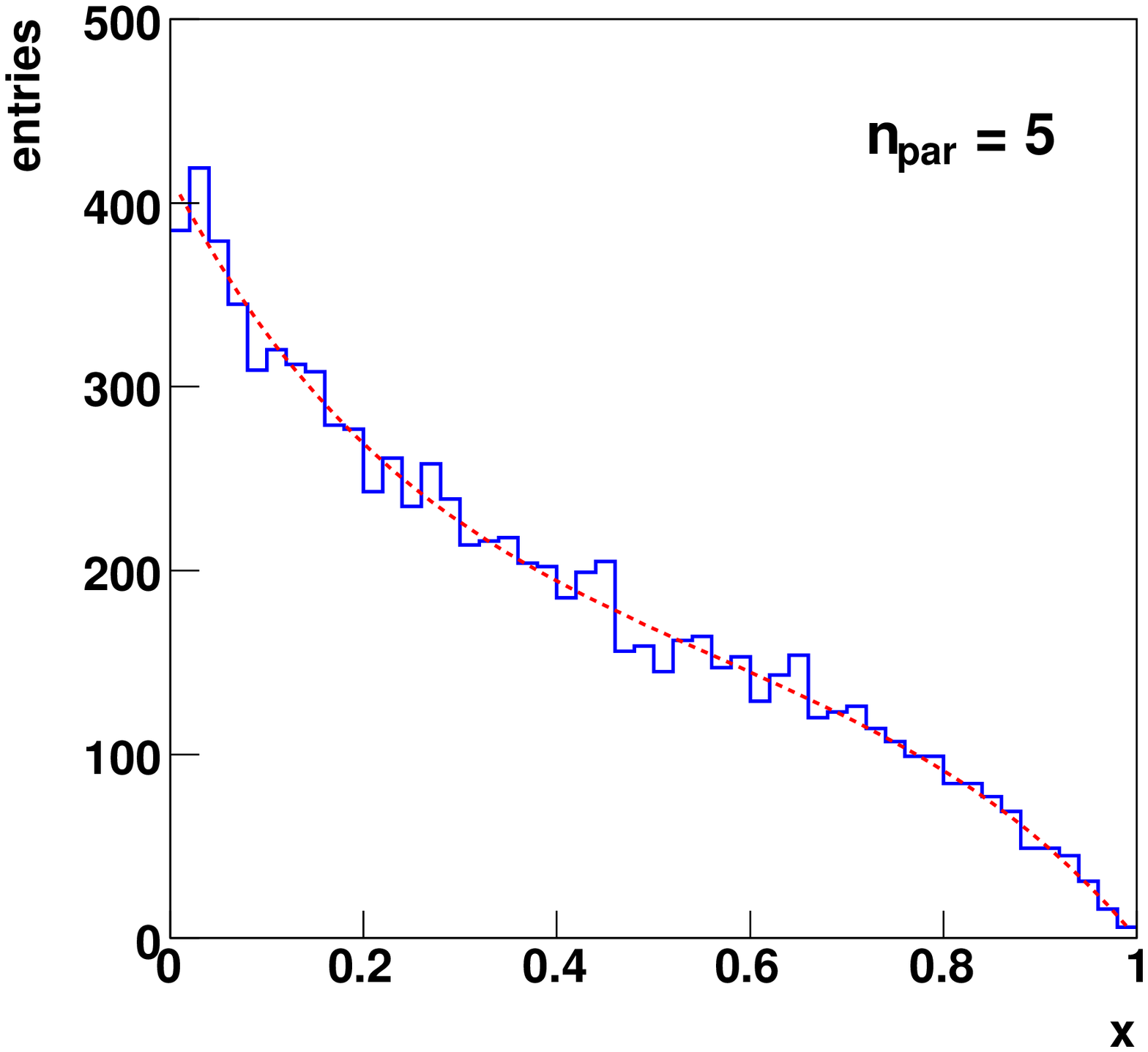}
}

\caption{ Fits for different numbers of adjustable parameters.
}\label{fig:fit}
\end{center}
\end{figure}

In the example shown in figure~\ref{fig:dist}, the true model
was in fact produced by distorting the distribution on the right with
a second order Bernstein polynomial using $\beta_0 = 1$, $\beta_1 = 0.5$
and $\beta_2 = 1.5$.  Recall that the original undistorted
distribution corresponds to $\beta_0 = \beta_1 = \beta_2 = 1$.  So in
this case one expects that to be well described the data will on
average require a scale factor based on a second-order basis function.
This is in fact what one sees in the fits shown in figure~\ref{fig:fit};
the fit with three adjustable parameters provides a good description
of the data.  Increasing the number of parameters further does not
improve substantially the goodness-of-fit.  This is also evident from
the values of the test statistic $q_{\vec{\nu}}$ and $q_{m,m+1}$ obtained
using different numbers of free parameters as shown in
table~\ref{tab:chi2}.

\setlength{\tabcolsep}{3.8mm}
\begin{table}[hbt]
\begin{center}
\caption{ Values of the variables $q_{\vec{\nu}}$ and $q_{m,m+1}$ 
and the corresponding $p$-values obtained from
fits with different numbers of adjustable parameters.
}\label{tab:chi2}
\begin{tabular}{|c|cc|cc|}
\hline
$n_{\rm par}$ & $q_{\vec{\nu}}$ & $p(q_{\vec{\nu}})$ & $q_{m,m+1}$ & $p(q_{m,m+1})$ \\
\hline
0  &  258.8  &  $6.1 \times 10^{-30}$  &  98.9  &  $2.6 \times 10^{-23}$  \\
1  &  159.9  &  $1.1 \times 10^{-13}$  &  15.4  &  $8.9 \times 10^{-05}$  \\
2  &  144.5  &  $1.3 \times 10^{-11}$  &  112.0  &  $3.5 \times 10^{-26}$  \\
3  &  32.5  &  0.95  &  0.0013  &  0.97  \\
4  &  32.5  &  0.93  &  0.26    &  0.61  \\
5  &  32.2  &  0.92  &  0.37    &  0.54  \\
\hline
\end{tabular}
\end{center}
\end{table}

\begin{figure}[htbp]
\begin{center}

\subfigure{
\includegraphics[width= 
0.30\textwidth]{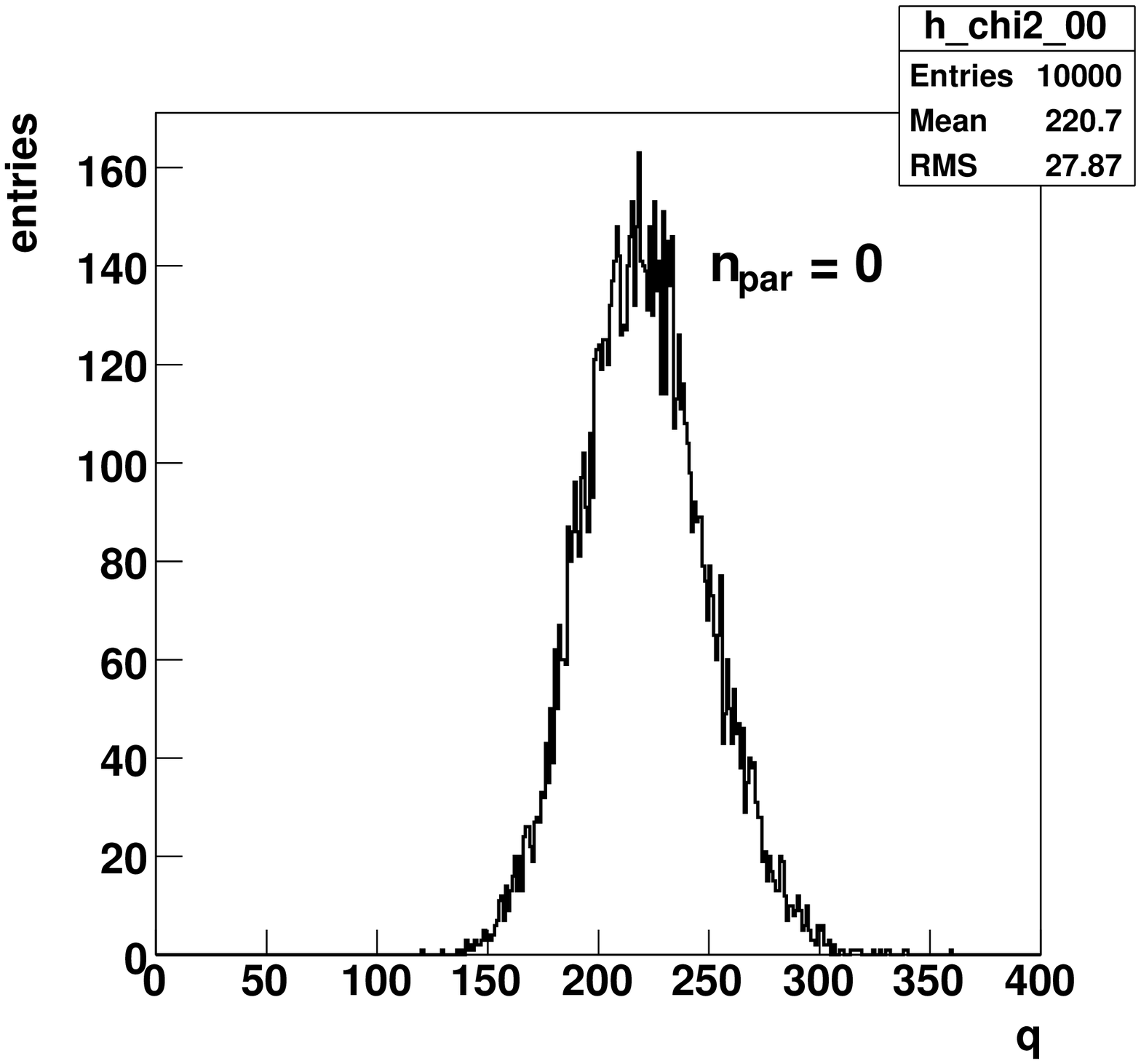}
}
\subfigure{
\includegraphics[width= 
0.30\textwidth]{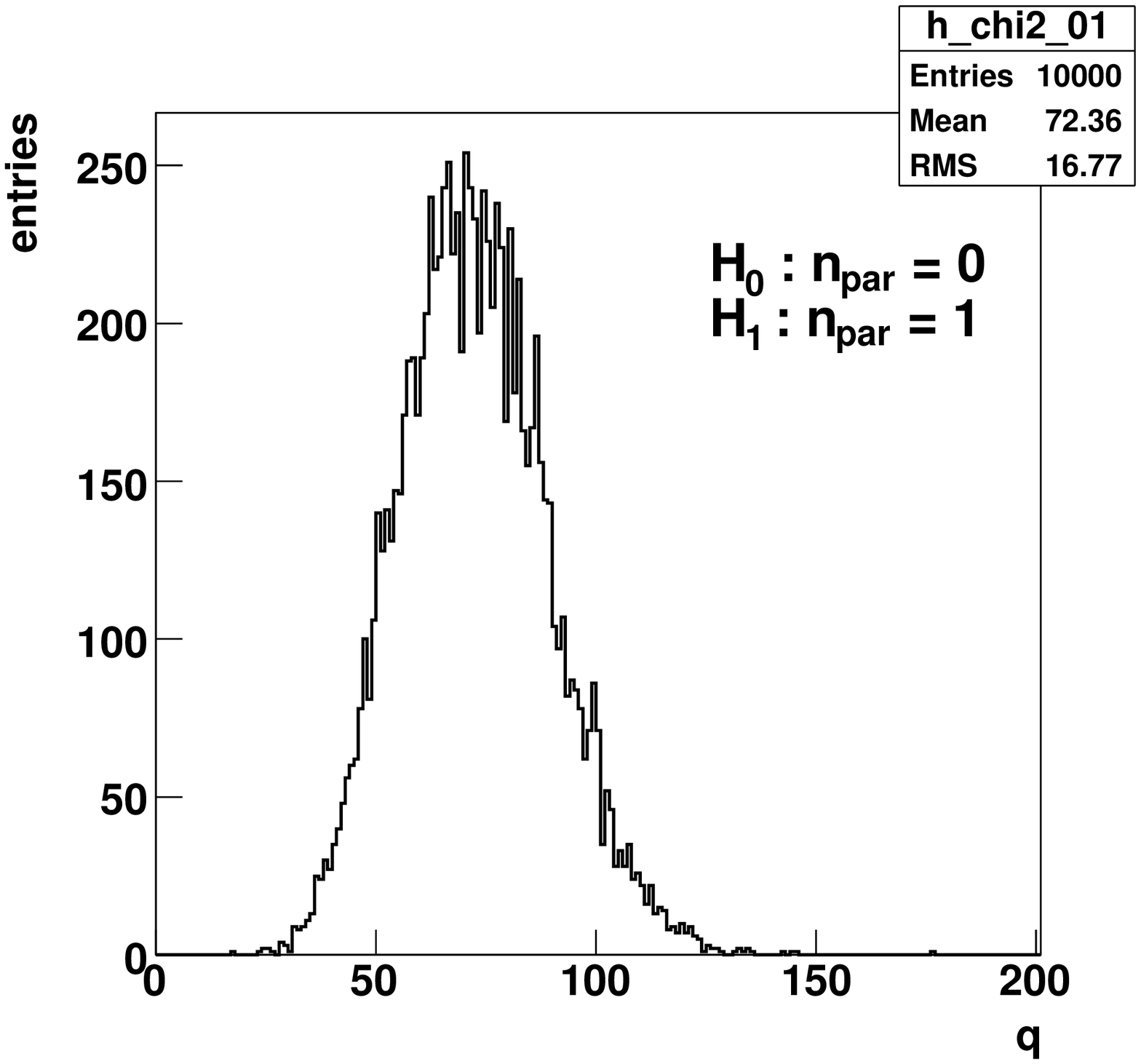}
}
\subfigure{
\includegraphics[width= 
0.30\textwidth]{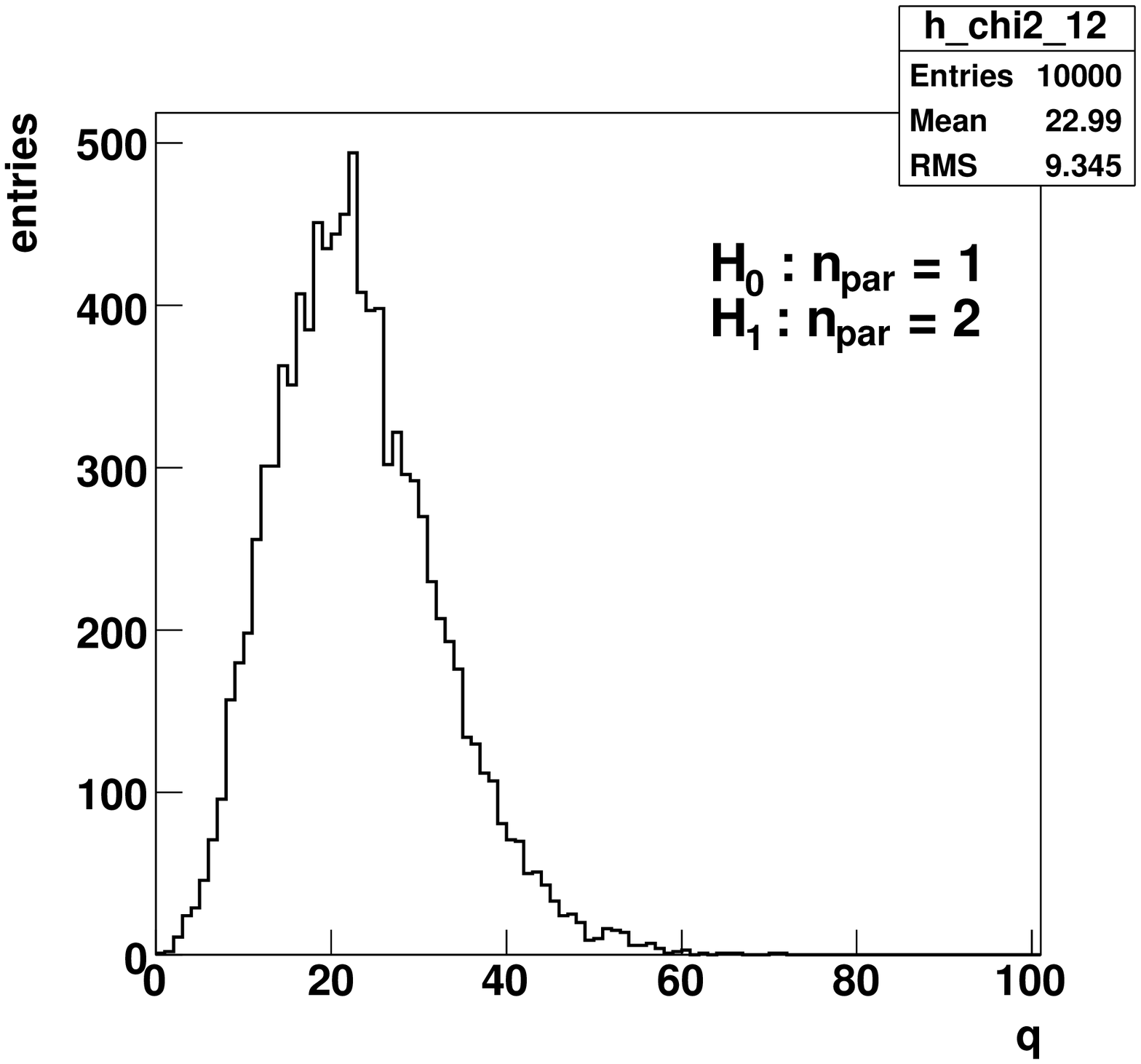}
}
\subfigure{
\includegraphics[width= 
0.30\textwidth]{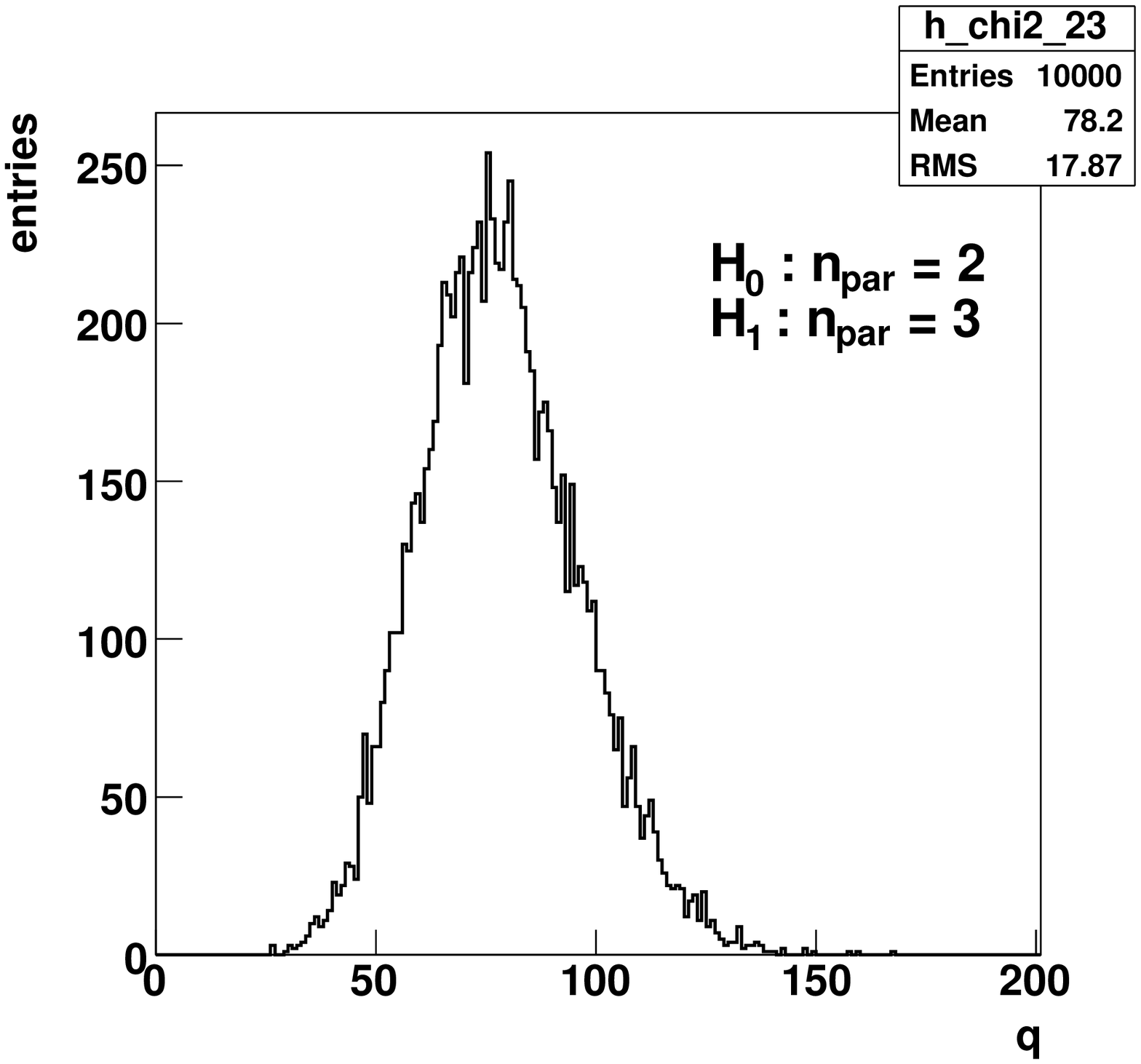}
}
\subfigure{
\includegraphics[width= 
0.30\textwidth]{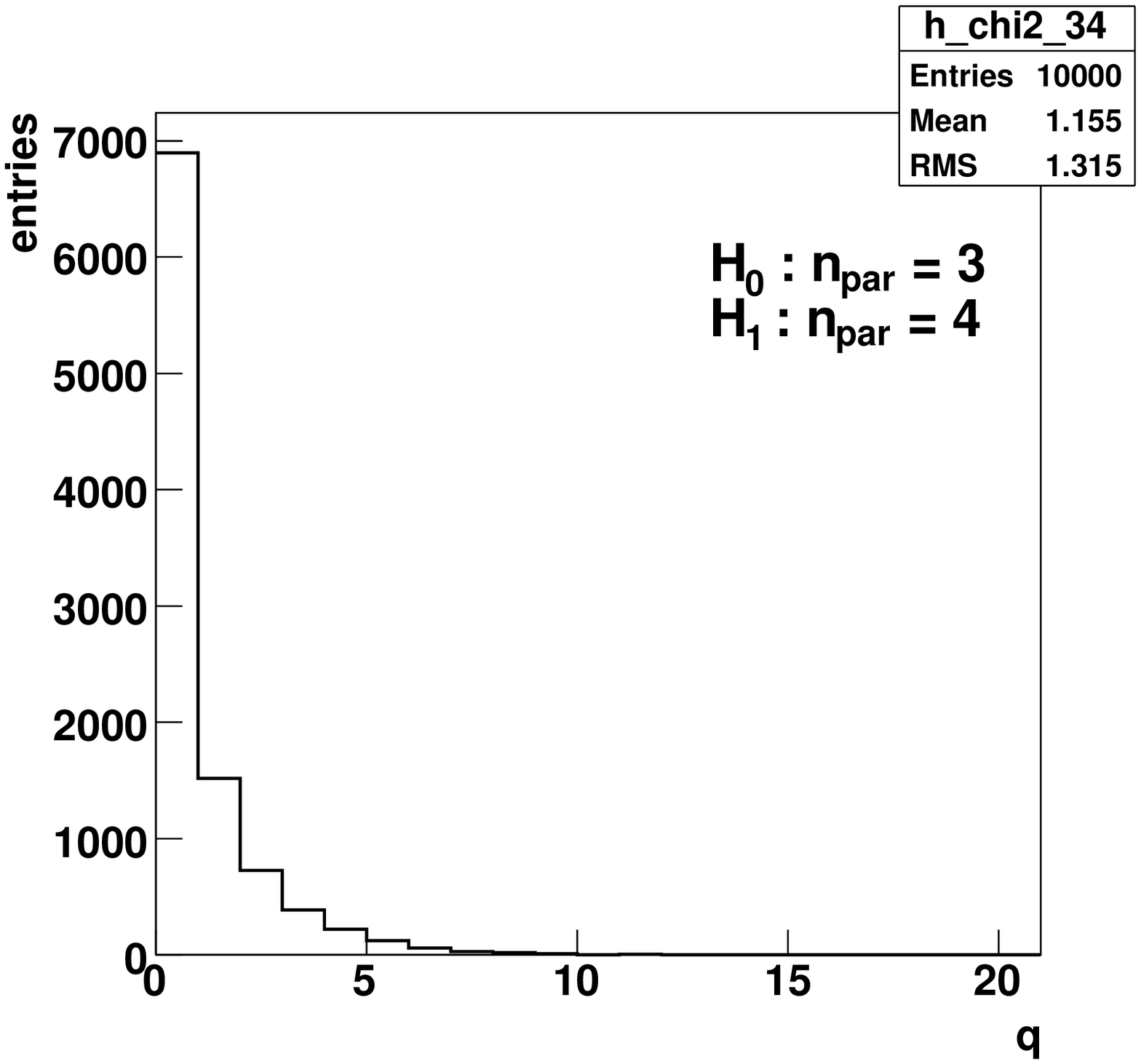}
}
\subfigure{
\includegraphics[width= 
0.30\textwidth]{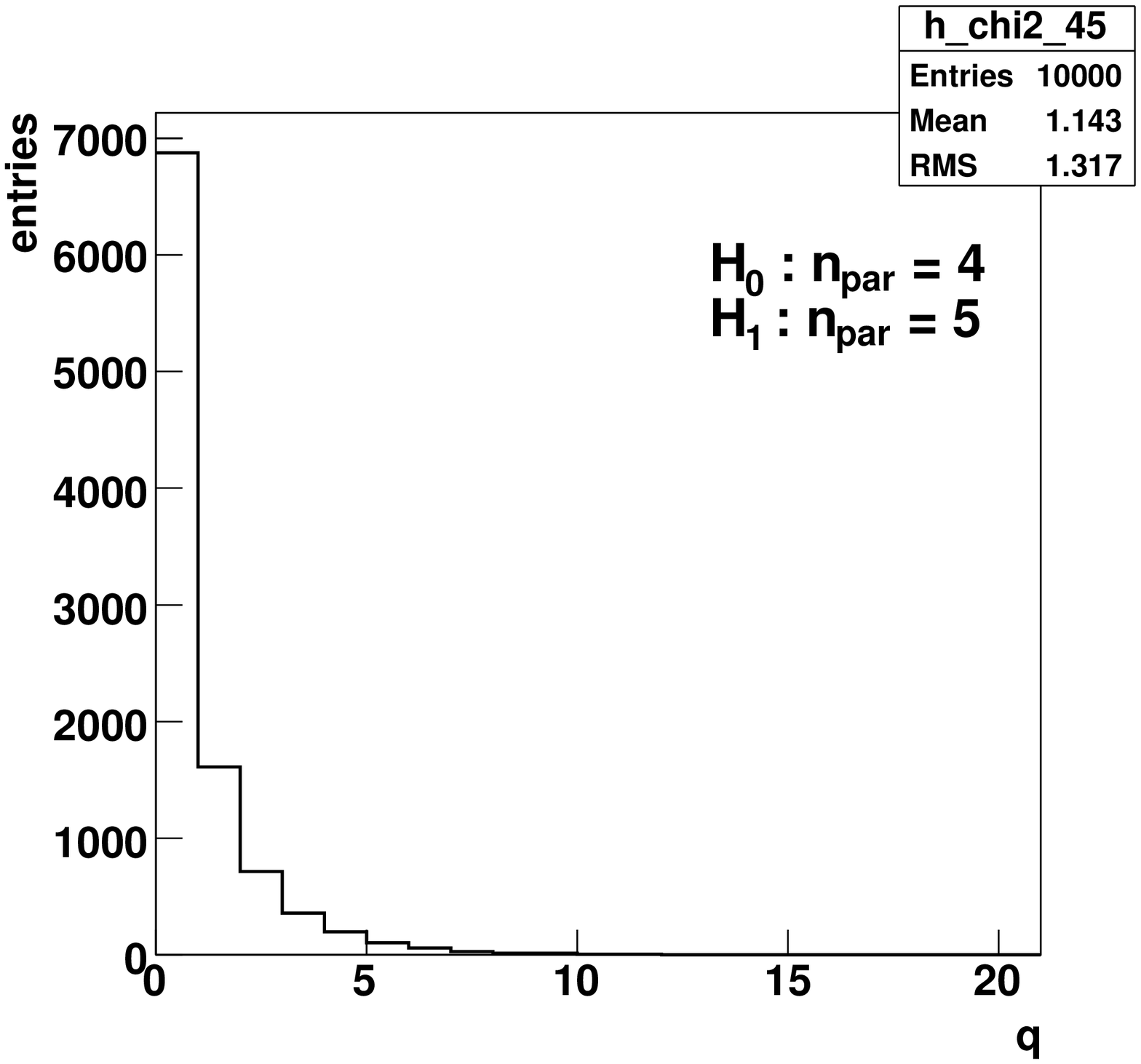}
}

\caption{ Distributions of the test variable $q$ (a)
for the zeroth-order model and (b)--(f) when comparing models
with differing numbers of additional parameters.
}\label{fig:chi2}
\end{center}
\end{figure}

We can investigate this further by simulating the experiment many times.
Figure \ref{fig:chi2} shows the distributions of the test variable $q$.
For figure \ref{fig:chi2}(a) this is the goodness-of-fit statistic
for the zeroth-order model.  For successive plots it shows the distribution
of $q_{m,m+1} = -2 \ln \lambda(m,m+1)$ where the likelihood ratio is based on the numbers
of parameters indicated.  From the plots one sees that distribution of $q_{m,m+1}$ 
based on the comparison of the 3 and 4 parameter models is close to a
chi-square distribution for one degree of freedom.  So in most cases one
would not reject the 3-parameter hypothesis.

\section{Performance and comparison with other background estimation techniques}

This section deals with the performance of the proposed method with respect to other common background estimation techniques in terms
of the total errors and the compensation of systematic effects.

\subsection{Data from control region as a model}

\begin{figure}[tbp]
  \begin{center}
    \includegraphics[width=.7\textwidth]{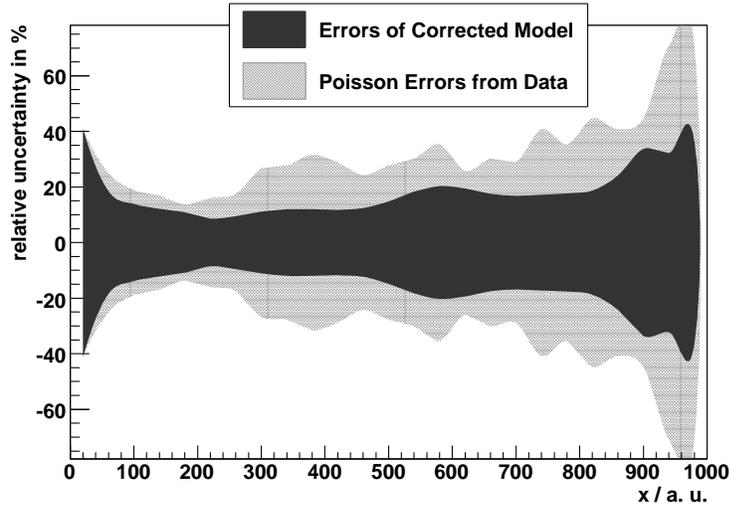}
  \end{center}
  \caption{ Relative total uncertainties of the data and the corrected background model as shown in figure  \protect\ref{fig:EstimatedAndTrueModelNOSys}.
  }\label{fig:ModelSelectionRelErrors}
\end{figure}

If a control region can be defined such that the shapes of the relevant background processes are practically identical to the
ones in the signal region a simple scaling of the data can be used to get a model for the background in the signal region. In
a first approximation, the uncertainties of such a model are simply the square root values of the data. For simplicity, assume
the efficiency of signal to control region to be unity. In this case the model determined in the control region can
be taken as-is for the signal region.  Consider again the second scenario discussed in subsection 3.1, depicted in figure \ref{fig:TwoScenarios}. Figure 
 \ref{fig:ModelSelectionRelErrors} compares the relative uncertainties. The corrected model represents a much smoother and thus more realistic
model than the data do -- see also figure \ref{fig:EstimatedAndTrueModelNOSys}. In addition, the uncertainties of the corrected model 
outperform bin-wise the Poisson errors of the data. Still, one has to account for the correlation in the first
case when summing events of several bins whereas the data are independently distributed.

\subsubsection{Background estimates for a new physics search}

\begin{figure}[tbp]
  \begin{center}
    \includegraphics[width=.7\textwidth]{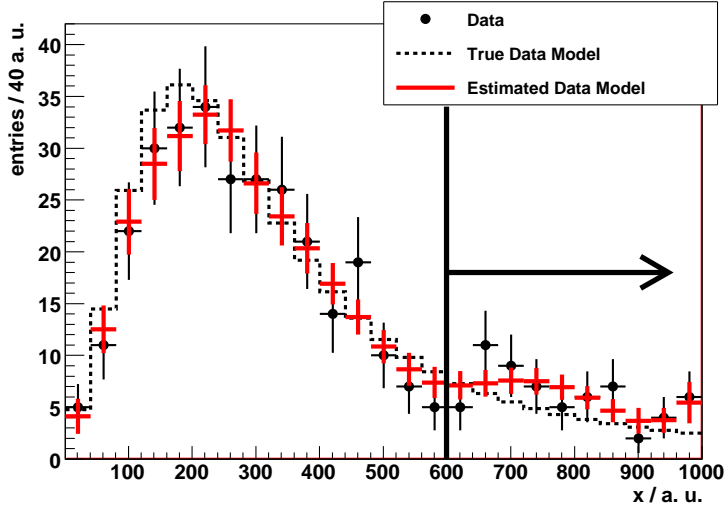}
  \end{center}
  \caption{ Summing events in the region $x > 600$ a.u., which is supposed to be sensitive for new physics.
  }\label{fig:SumEvents600}
\end{figure}

In a search for new physics one is often interested in the high mass tails of distributions for being the most sensitive regions to
discover new phenomena. Suppose this region to include all $x$-values greater than 600 a.u.~-- see figure \ref{fig:SumEvents600}. 
Table \ref{tab:EventsGreater600} summarizes the expected number of events and its uncertainty for the original prediction, the corrected 
model and when taking the data as the model. Both the data model and the corrected model have a comparable and much smaller uncertainty 
than the original prediction. The abovementioned correlation boosts the error of the corrected model to the level of the data uncertainty
in this example.

\begin{table}[tbp]
\begin{center}
	\caption{ Number of expected events for $x > 600$ a.u.~predicted by different models (cf.~figure \protect\ref{fig:SumEvents600}).
			The error of the corrected model is the same as the one from the data in this case, in general it is smaller
			(see figure \protect\ref{fig:ExpEvents600au}).
			The MC template is identical to the true data model since no systematic effects were introduced in this scenario.
	     }\label{tab:EventsGreater600}
	\vspace{5pt}
	\begin{tabular}{|l|c|c|} \hline
	Model & Number of expected events & Relative error\\
	\hline
	Original prediction (MC template) & $43.9  \pm  21.9$ & $50 \% $\\ 
	Corrected model & $59.9  \pm  7.6$ & $12.7 \% $ \\ 
	Data as model & $62.0  \pm  7.9$ & $12.7 \%$ \\ 
	\hline
	\end{tabular}
\end{center}
\end{table}

\begin{figure}[tbp]
\begin{center}

\subfigure{
\includegraphics[width= 
0.45\textwidth]{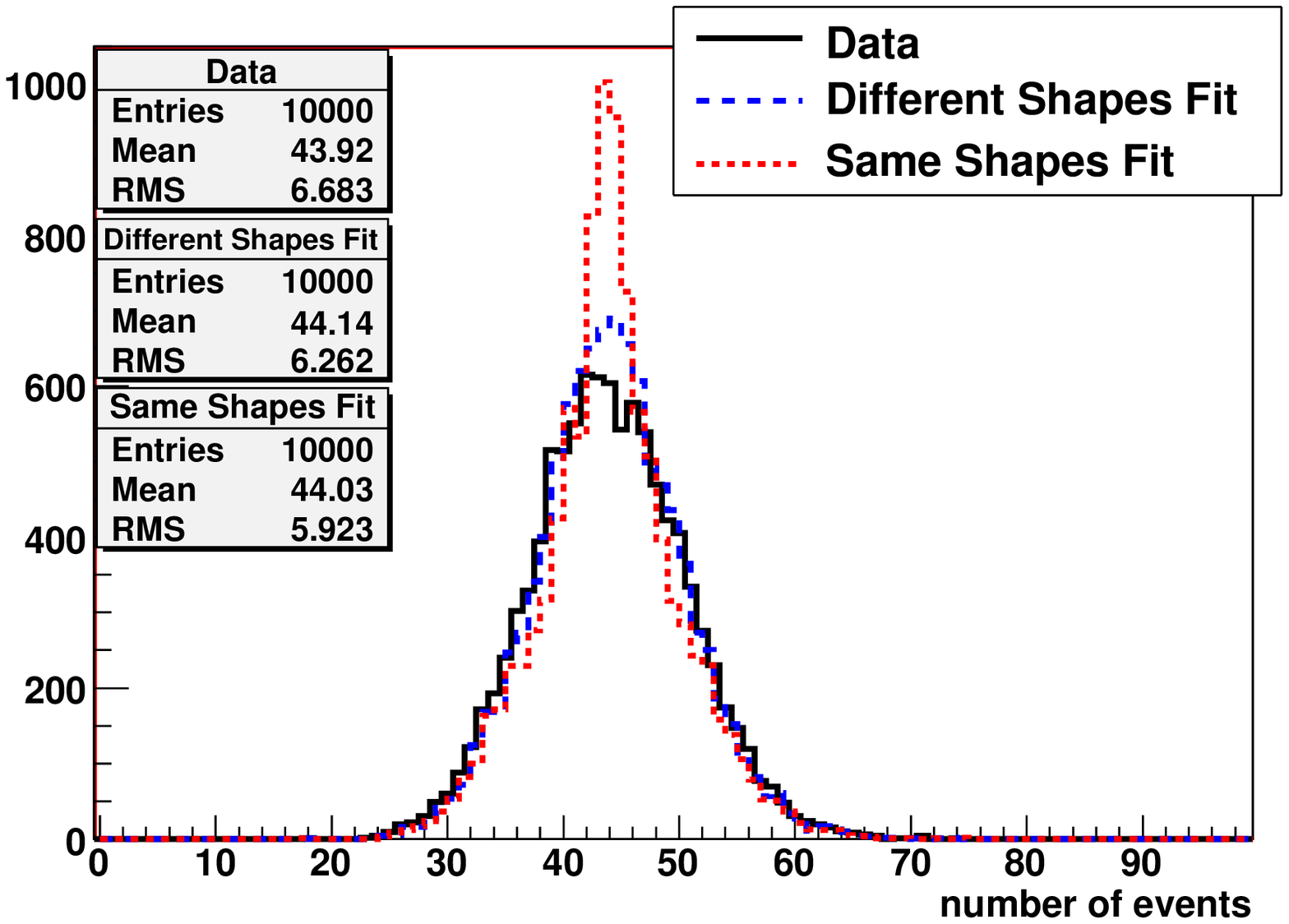}
}
\subfigure{
\includegraphics[width= 
0.45\textwidth]{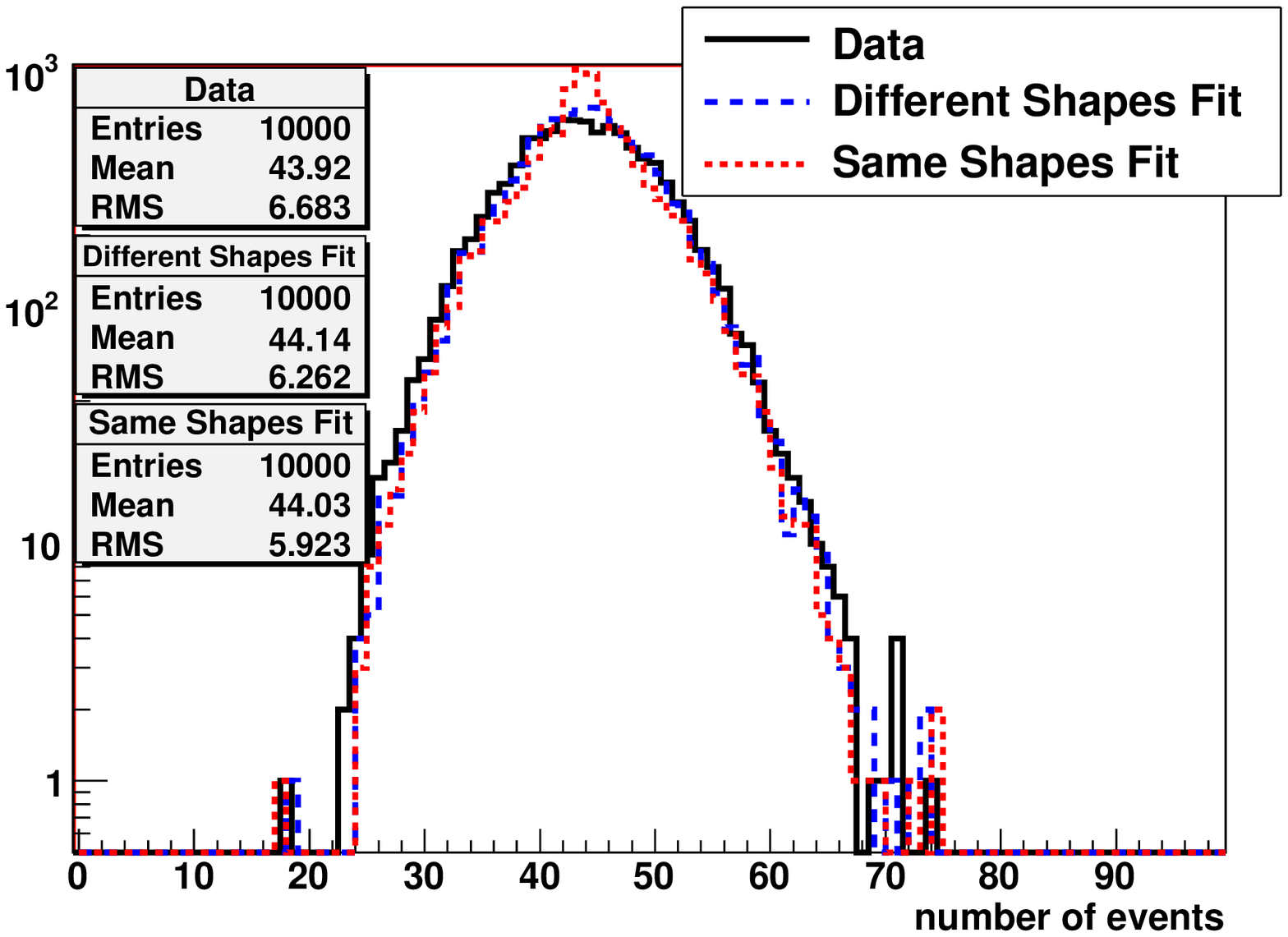}
}
\caption{ Distributions of number of expected events in the region $x>600$ a.u.~of figure \protect\ref{fig:SumEvents600} with linear and logarithmic
ordinate. The proposed method using the "different shapes" from figure \protect\ref{fig:VariousTemplates} yields a smaller RMS than using
the data as a model. The uncertainty can be further reduced by using templates more similar to the true model, in this case only differing
in the scale but having the "same shape".
}\label{fig:ExpEvents600au}
\end{center}
\end{figure}

In order to obtain a general statement on how the error of the proposed method compares with the one from the data 10000 pseudo data
sets are created from the true data model. Figure \ref{fig:ExpEvents600au} shows the distribution of events for x-values greater than
600 a.u. Taking the data as the model, it produces an unbiased prediction of 43.92 events for the mean value with an error of 6.68 events, as 
expected in agreement with the true values of 43.89 and 6.63 within the statistical limitation of the sample. Applying the proposed 
method yields on average a value of 44.14 and a reduced error of 6.26. The mean value is slightly positively biased but only by about 
4\% of the quoted uncertainty. This bias originates from fits with small $p$-values in the percent regime. It can be reduced by either
simply vetoing such fits with highest absolute $p$-values less than for instance 1\% or by allowing fits with more than 10 parameters.

More knowledge about the true shape of the distribution can reduce the uncertainty of the method even further since fewer parameters
will be needed for the adjustment of appropriate starting templates. As a limiting case, five templates which only differ in their 
normalization with respect to the true model are employed. The resulting distribution of expected events is also displayed in figure 
\ref{fig:ExpEvents600au} as the dashed red line, demonstrating a further decrease of the error to 5.92.

\subsubsection{Significance of a possible signal}

When calculating a significance of a number of observed events given the Standard Model expectation, one
can include the uncertainty on the latter by using a Bayesian prior. Since the uncertainty often arises from
various sources one can assume a Gaussian distribution of the prior as stated by the central limit theorem
of probability theory. Therefore, it is desirable for the error of the method to exhibit a Gaussian behaviour.

\begin{figure}[tbp]
\begin{center}

\subfigure{
\includegraphics[width= 
0.45\textwidth]{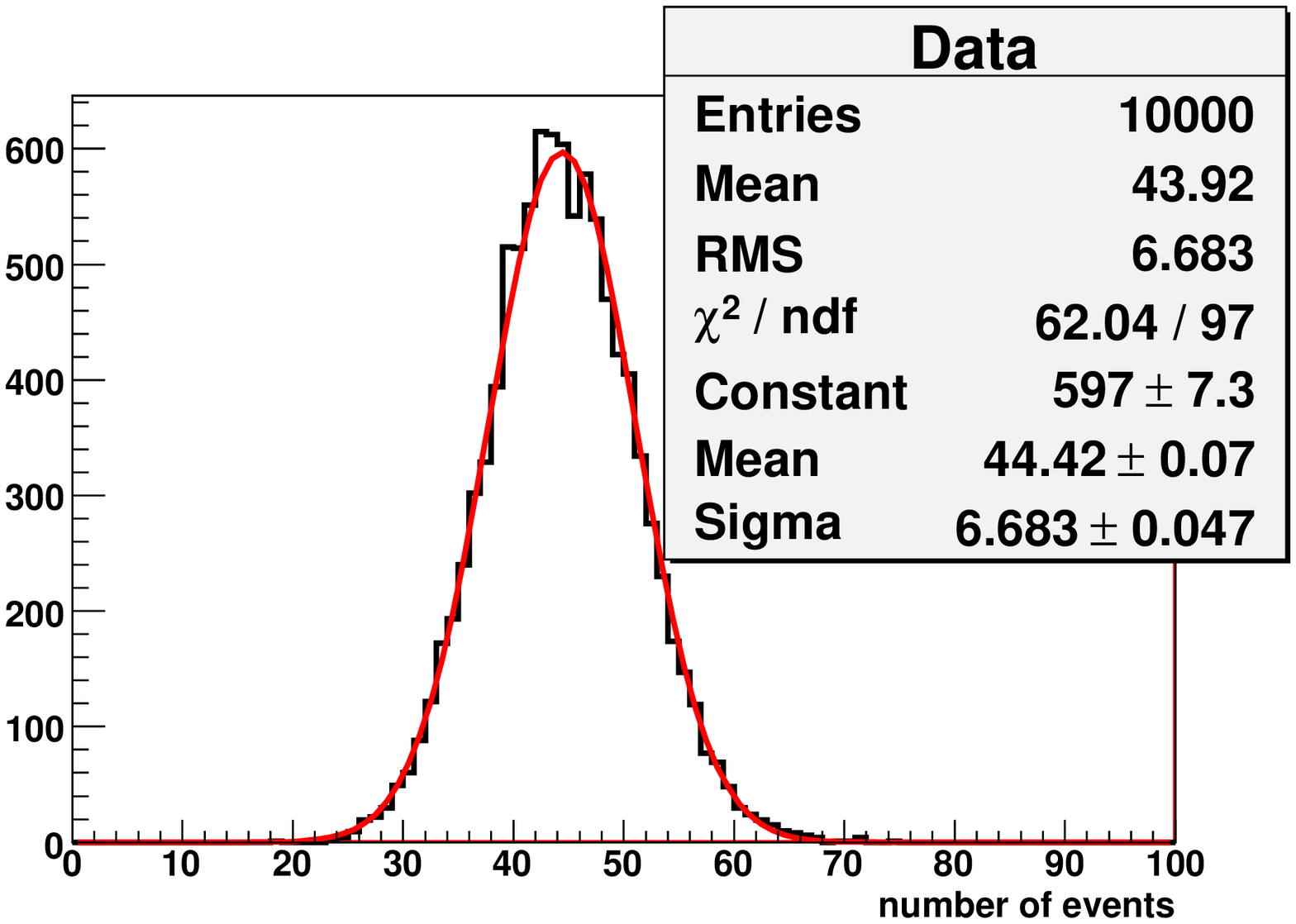}
}
\subfigure{
\includegraphics[width= 
0.45\textwidth]{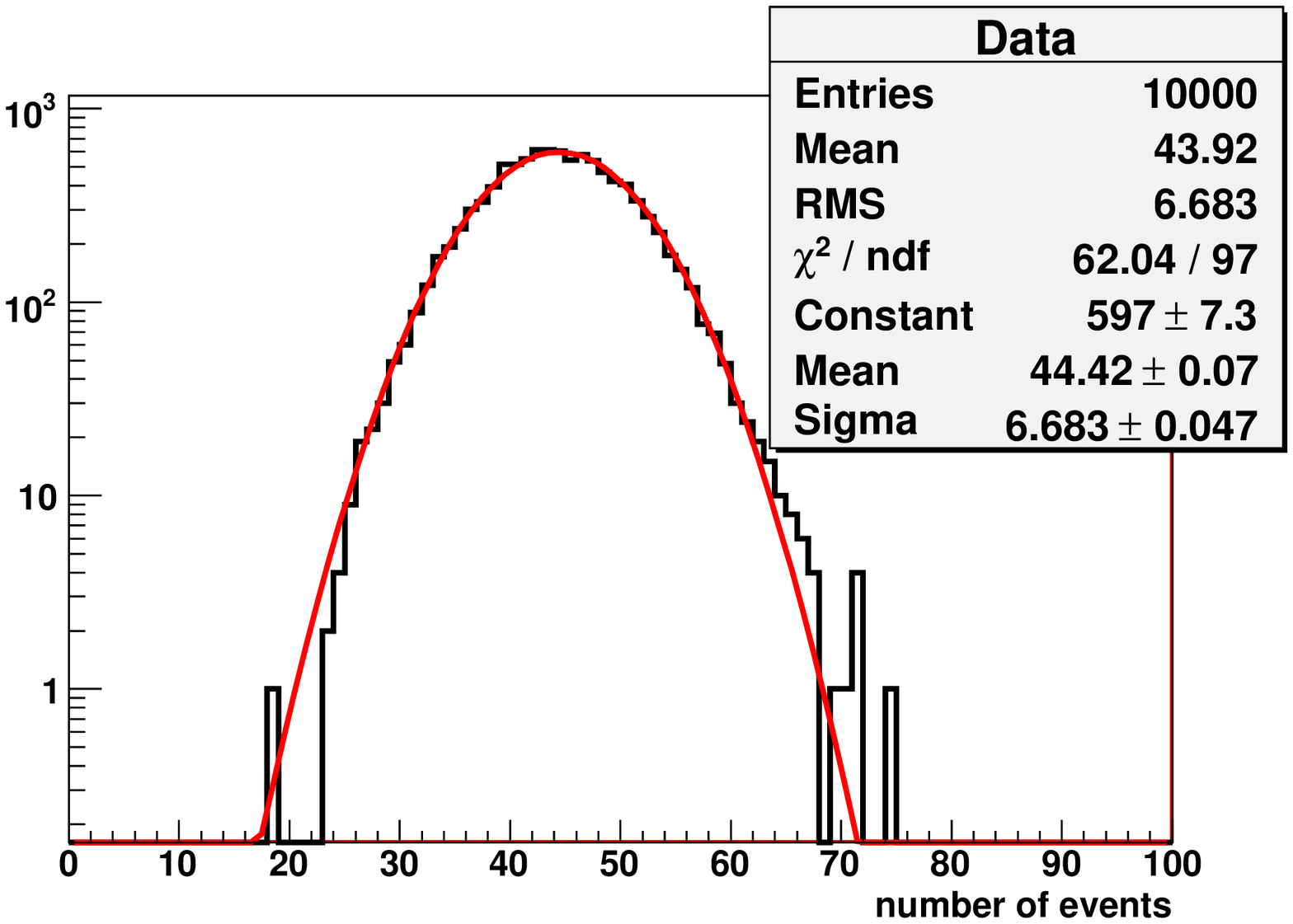}
}

\subfigure{
\includegraphics[width= 
0.45\textwidth]{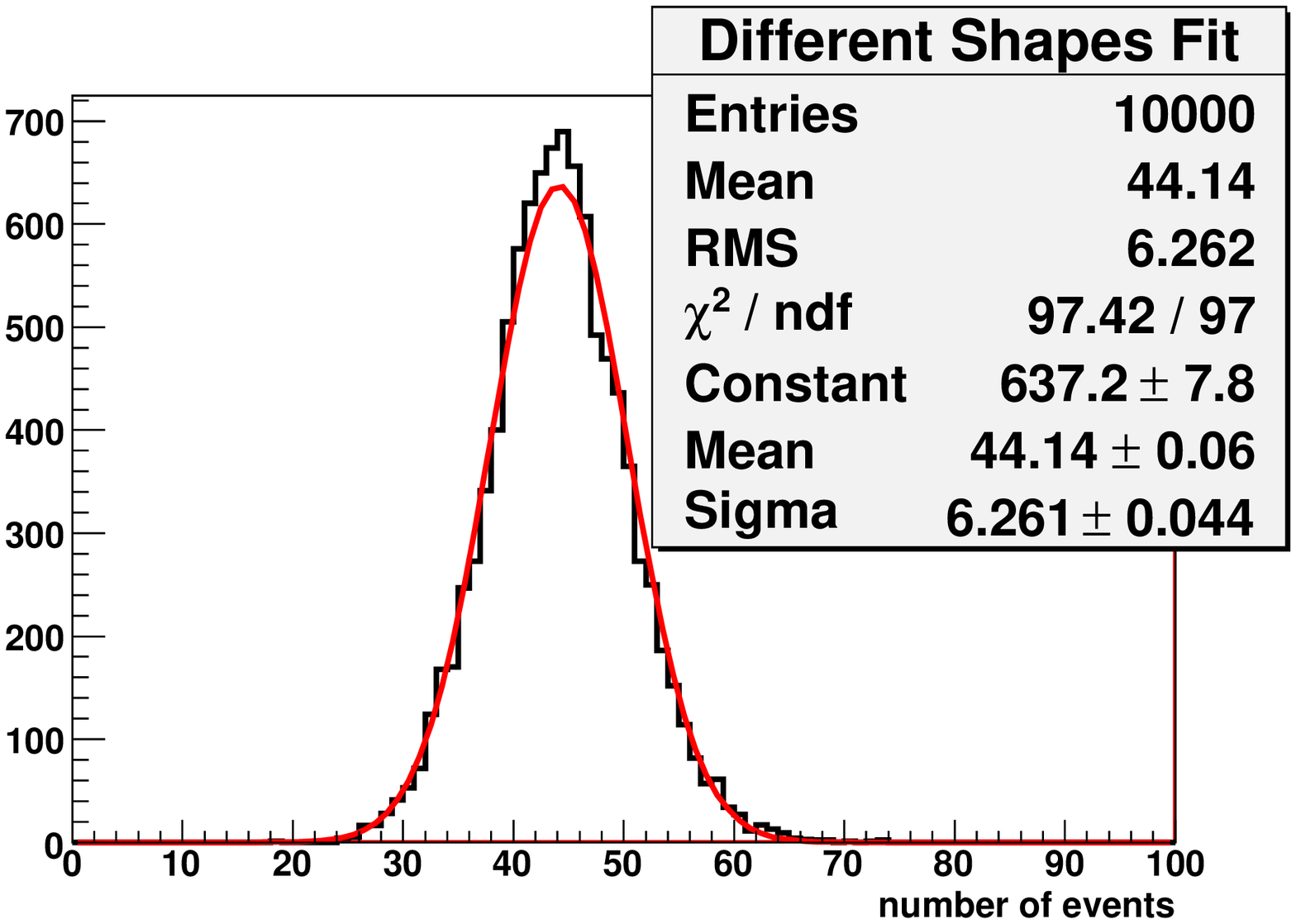}
}
\subfigure{
\includegraphics[width= 
0.45\textwidth]{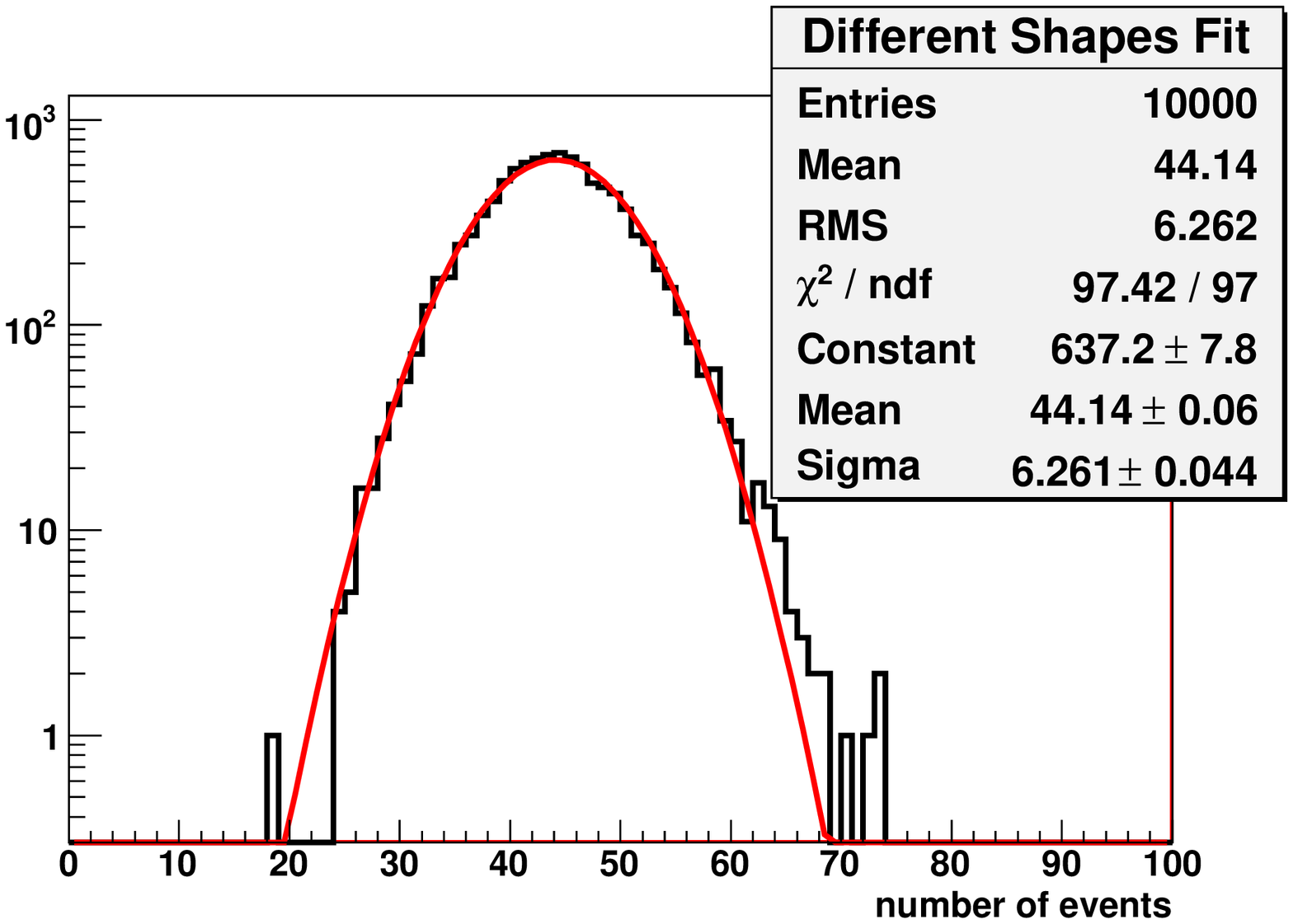}
}

\subfigure{
\includegraphics[width= 
0.45\textwidth]{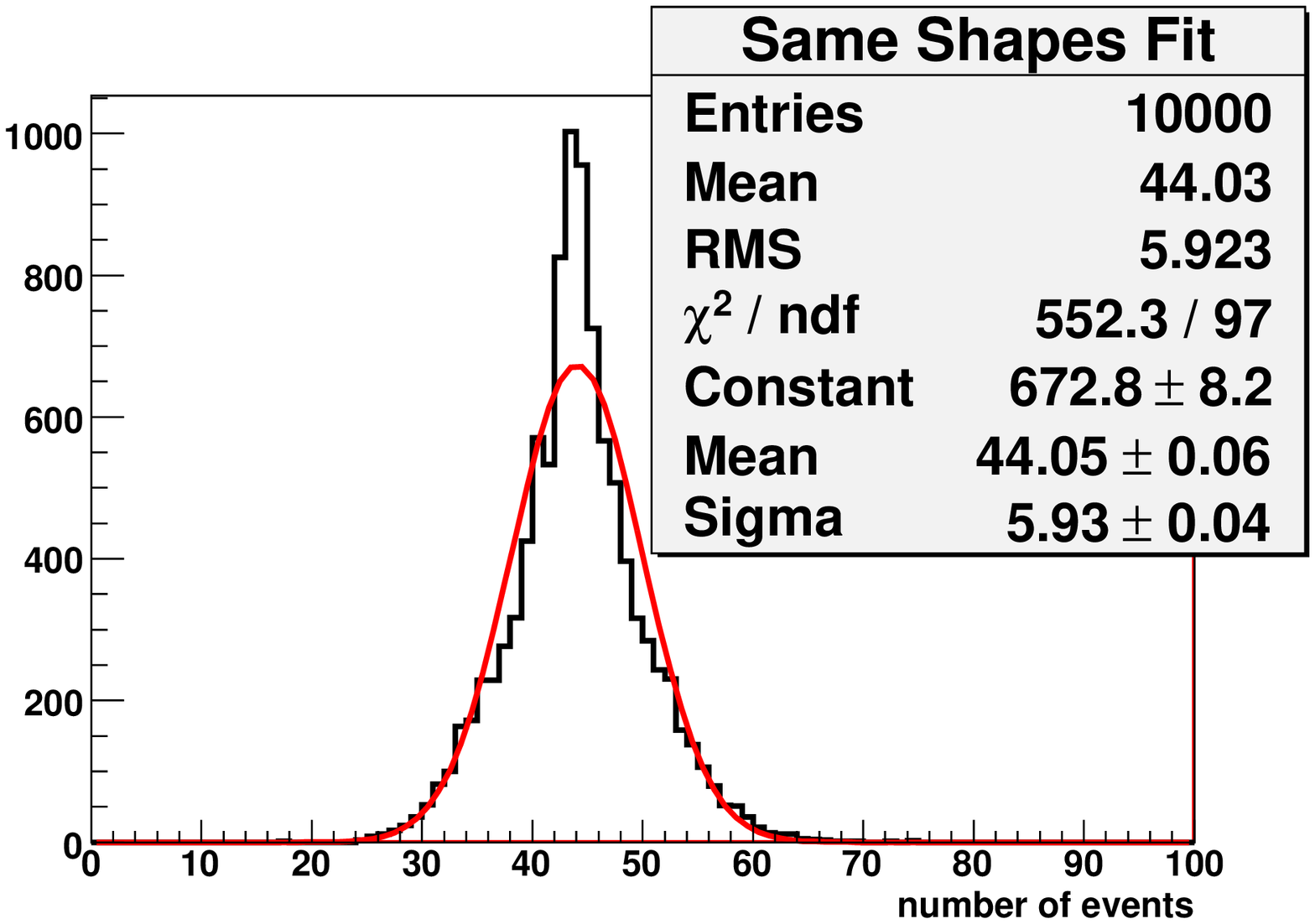}
}
\subfigure{
\includegraphics[width= 
0.45\textwidth]{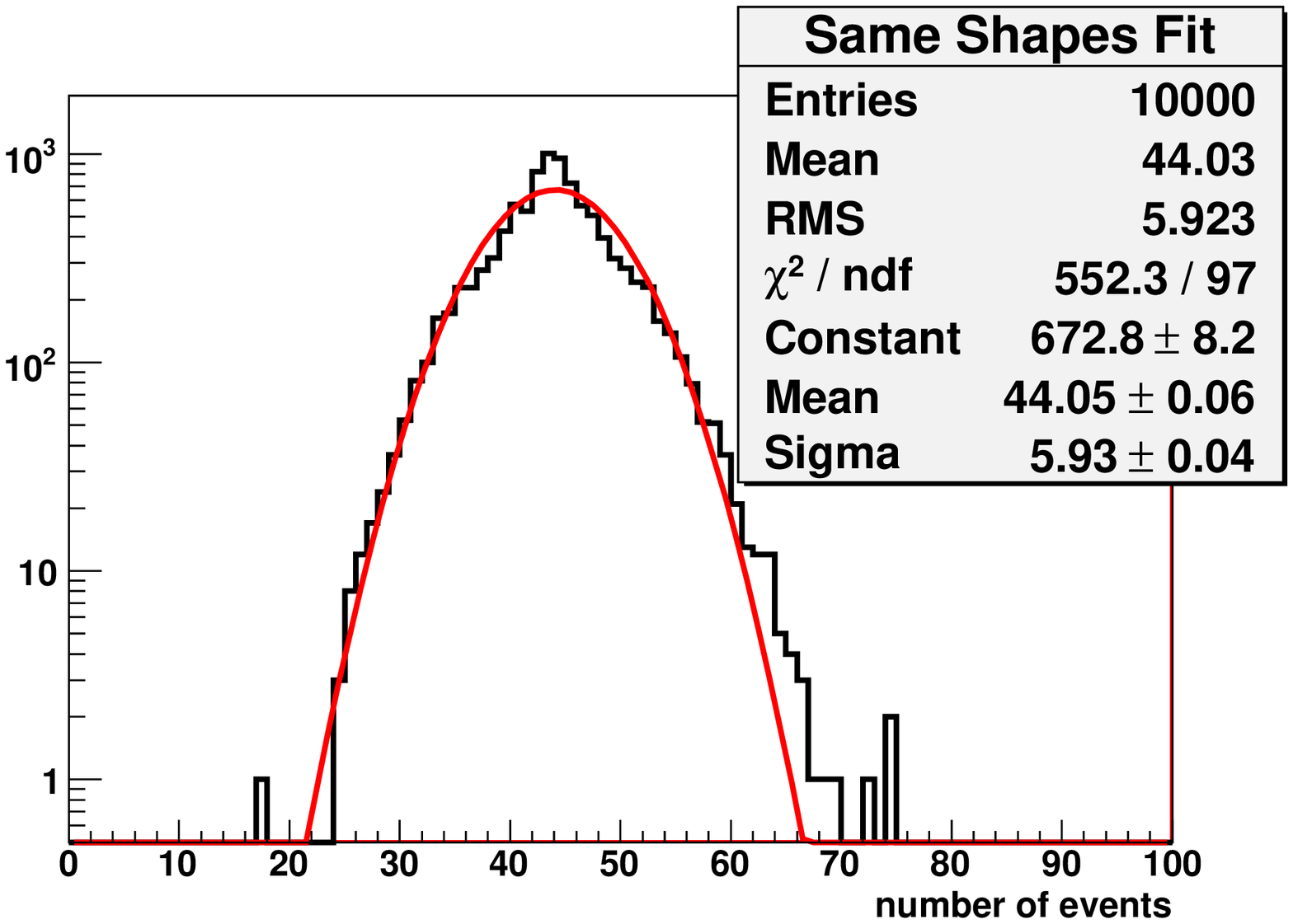}
}
\caption{ Gaussian functions fitted to the distributions of figure \protect\ref{fig:ExpEvents600au}. Both the "data" and the
"different shapes fit" display an acceptable Gaussian behaviour whereas the "same shapes fit" does not, but still does produce
the same RMS value.
}\label{fig:ExpEvents600auGaussianFits}
\end{center}
\end{figure}

The six plots of figure \ref{fig:ExpEvents600auGaussianFits} show the number of expected events for the region
$x>600$ a.u.~fitted with a Gaussian function both with linear and logarithmic ordinate for the three different
predictions from figure \ref{fig:ExpEvents600au}. As expected the data display an exact Poisson distribution which
can be well approximated with a Gaussian for high enough mean values. Also the uncertainty of the proposed method
using the different shapes from figure \ref{fig:VariousTemplates} is compatible with a normal approximation except
for acceptable deviations in the central and tail regions. Using templates only differing in scale from the true
background model entails an uncertainty which varies from the Gaussian fit. However, the RMS value of 5.92 events 
agrees nicely with the width of the fitted Gaussian. 
Furthermore, the Gaussian behavior is expected to improve when accounting for additional systematic
effects associated with the transfer of the background model from the control to the signal region.

In order to investigate how the different errors affect the discovery potential, two toy measurements for the two regions $x>600$ a.u.~and
 $x>800$ a.u.~are assumed to be 99 and 52 events respectively as shown in table \ref{tab:SignificancesEventsGreater600}. High energy
physics folklore considers a measurements to be a discovery if the probability, assuming only known physics, of observing data as or less
likely is smaller than $2.9 * 10^{-7}$, which corresponds to the integrated tail of a Gaussian distribution beyond five standard deviations
("$5 \sigma$ discovery'').

Using the data from the control region as the background model one would claim a discovery since the significance, which is calculated by convoluting the Poisson probability for the data with the Gaussian prior function representing the systematic uncertainty of the background (see e.g.~\cite{Linnemann} and
\cite{CSC}),
surpasses the $5 \sigma$ threshold. Taking instead the predicted mean value and error of the proposed method using the different starting templates
the significance grows to 5.12 and 5.29 for the two regions. It can be even further raised to 5.25 and 5.38 when using the set of "same shape templates". 
The jump in significance is equivalent to an increase in luminosity of 4\% and 12\% for the two regions respectively when using the same shape
fit results instead of the data. Thus, by using the proposed method the required integrated luminosity for a discovery is reduced. This effect gets bigger
the smaller the inspected tail region compared to the region in $x$ which was used to determine the background model.

\begin{table}[tbp]
\begin{center}
	\caption{ The significance of a discovery can be increased by using the proposed method instead of data from the control region
		as a background model. The measurements of 99 and 52 events for the two regions were chosen to allow for a $5 \sigma$ discovery
		when using the data. The increase in significance is equivalent to a saving in luminosity as described in the text. 
	     }\label{tab:SignificancesEventsGreater600}
	\vspace{5pt}
	\begin{tabular}{|l|cc|cc|} \cline{2-5}
	\multicolumn{1}{c|}{}& \multicolumn{2}{c|}{$x>600$ a.u.: 99 events}  & \multicolumn{2}{c|}{$x>800$ a.u.: 52 events}\\
	\multicolumn{1}{c|}{}& Background: & Significance: & Background: & Significance:	\\
	\hline
	Data & $43.92  \pm  6.68$ & 5.01 & $15.62 \pm 3.93$  & 5.10\\ 
	Different Shapes & $44.14  \pm  6.26$ & 5.12 & $15.56 \pm 3.60$  & 5.30  \\ 
	Same Shapes & $44.03  \pm  5.92$ & 5.25 & $15.53 \pm 3.45$  & 5.38 \\ 
	\hline
	\end{tabular}
\end{center}
\end{table}

\subsection{Parametrized Monte Carlo shapes}

A different approach to estimate the background in the signal region lies in choosing appropriate parametric functions inspired by the shapes of the Monte Carlo estimates and fixing the parameters through a fit to data in the control region. These predictions can then be extrapolated to the signal region by scaling.

\begin{figure}[tbp]
  \begin{center}
    \includegraphics[width=.7\textwidth]{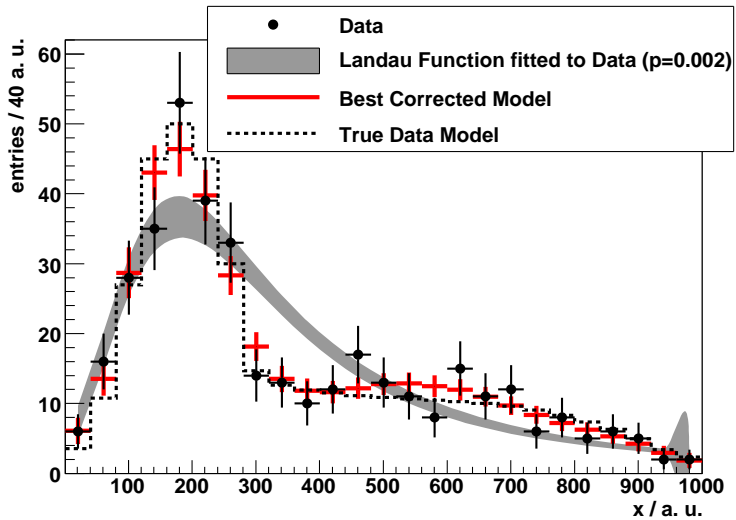}
  \end{center}
  \caption{ Data (black) in control region (cf.~figure \protect\ref{fig:ModelSelectionPlot}) fitted with a Landau function (grey band), the choice of which was
	 inspired by the Monte Carlo estimate (dashed line in figure \protect\ref{fig:ModelSelectionPlot}). The systematic effects cannot be compensated by an
	 adjustment of the Landau function which is reflected by a poor $p$-value of about 0.2\%. 
  }\label{fig:NikhefCR}
\end{figure}

Examining the Monte Carlo template in figure \ref{fig:TwoScenarios} one could guess that a Landau function, which depends on three parameters, might describe the background sufficiently. This is the case because the MC template is a discretized Landau function. The outcome of the fit to data of the first scenario is shown in figure \ref{fig:NikhefCR}. While having a much smaller uncertainty than the background data scaling method the predicted shape is incompatible with the data in this example as reflected by a $p$-value $p(q_{\vec{\nu}})=0.002$. Obviously, if the systematic effects can't be compensated by an adjustment of the function's parameters, the estimate will differ significantly from the true background. 

\subsection{Direct fit to data in control region}
Alternatively, one could fit the data in the control region for the first scenario with a suitable polynomial function to compensate for the 
strong systematic effects, thereby dismissing any prior knowledge about the model. In order to determine the best-suited polynomial the same 
statistical test used for the proposed method ($p(q_{\vec{\nu}})$) was employed, selecting a function with 9 parameters. Table 
\ref{tab:PolynomialFit} shows the expected number of events predicted by the different approaches. The polynomial fit produces a worse result 
than the proposed method, which yields identical values to the data in this case, both with respect to the predicted value and its uncertainty. 
So even in this case, where a lot of parameters are required to adjust the templates to the true model, the proposed method outperforms a 
direct polynomial fit.

\begin{table}[tbp]
\begin{center}
	\caption{ Number of expected events predicted by different models for the scenario with large systematic effects
		(cf.~figure \protect\ref{fig:EstimatedAndTrueModel}). The errors of the corrected model and the polynomial fit 
		were calculated using 10000 pseudo data sets. The value of the true model amounts to 376.7 events.
	     }\label{tab:PolynomialFit}
	\vspace{5pt}
	\begin{tabular}{|l|c|c|} \hline
	Model & Number of expected events & Relative error\\
	\hline
	Original prediction (MC template) & $352.9  \pm  176.5$ & $50 \% $\\ 
	Data as model & $380.0  \pm  19.5$ & $5.1 \%$ \\
	Corrected model & $380.0  \pm  19.5$ & $5.1 \% $ \\ 
	Polynomial fit of order 8  & $363.6  \pm  21.2$ & $5.8 \% $ \\
	\hline
	\end{tabular}
\end{center}
\end{table}

\section{Summary and conclusion}

The underlying idea of the method presented in this paper is to correct the Monte Carlo background estimates for systematic deviations. To that end, they are multiplied with successively more complex correction functions until a statistical test reports good compatibility with data in a control region. The correction determined that way is then applied on the corresponding templates in the signal region yielding an improved background model to search for new physics. 

While systematic effects are absorbed by the correction functions, the total uncertainty of the model can be reduced compared to other common methods.
In order to avoid absorbing a possible signal in the fit carried out in the control region the Monte Carlo estimates can be varied according to known systematic effects, thereby obtaining constraints on the maximal acceptable modification of the templates.

Finally, the usefulness of the proposed method is not restricted to high energy physics. It can be applied in other scientific fields where
one uses data from control regions to estimate the background in a signal region and is confronted with large systematic uncertainties.

\acknowledgments
S.~Caron, S.~Horner and J.~E.~Sundermann gratefully acknowledge the support by the BMBF, the Landesstiftung Baden-W\"urttemberg 
and the Graduiertenkolleg ''Physik an Hadron-Beschleunigern'' of the DFG.
E.~Gross is obliged to the Benoziyo center for High Energy
Physics, to the the Israeli Science Foundation (ISF), the Minerva Gesellschaft
and the German Israeli Foundation (GIF) for supporting this work.


\begin{thebibliography}{[9]}

\bibitem{Wilks} S.S.~Wilks, {\emph{The large-sample distribution of the
likelihood ratio for testing composite hypotheses}}, \href{http://projecteuclid.org/euclid.aoms/1177732360}{\emph{Ann.\ Math.\
Statist.}\ {\bf 9} (1938) 60}.

\bibitem{Kendall2}  A.~Stuart, J.K.~Ord, and S.~Arnold,
{\it Kendall's advanced theory of statistics. Volume 2A: classical
inference and the linear model}, 6$^{\textnormal{th}}$ edition, Oxford University Press, Oxford U.K. (1999),
and earlier editions by Kendall and Stuart.

\bibitem{Bernstein} Wikipedia contributors, \emph{Bernstein polynomial} in \emph{Wikipedia, the free encyclopedia},
online at
\href{http://en.wikipedia.org/wiki/Bernstein\_polynomial}{http://en.wikipedia.org/wiki/Bernstein\_polynomial}, accessed December 15 (2008).


\bibitem{Linnemann} J. T. Linnemann, {\emph{Measures of significance in HEP and astrophysics}},
in the proceedings of \emph{PhyStat2003: Statistical Problems in Particle Physics,
Astrophysics, and Cosmology}, September 8-11, SLAC, Stanford, California U.S.A. (2003) [\physics{0312059}], see online at
\href{http://www.slac.stanford.edu/econf/C030908/papers/MOBT001.pdf}{http://www.slac.stanford.edu/econf/C030908/papers/MOBT001.pdf.} 

\bibitem{CSC} ATLAS collaboration, G. Aad et al.,
 {\emph{Expected performance of the ATLAS experiment,
 detector, trigger and physics} }, [\href{http://www.arxiv.org/abs/0901.0512}{\tt arXiv:0901.0512}].

\end{thebibliography}
\end{document}